\documentclass{LMCS}

\usepackage{enumerate}
\usepackage{hyperref}

\usepackage{proof}
\usepackage{latexsym}
\usepackage{amssymb}

\newcommand\cred[1]{\stackrel{#1}{\leadsto}}
\newcommand\credtr[1]{\stackrel{#1}{\Longrightarrow}}
\def\credid{\cred {}}

\newcommand\enc[2]{\{#1\}_{#2}}
\newcommand\dec[2]{{\mathrm{dec}}(#1,#2)}

\newcommand\spr[2]{\langle #1, #2 \rangle}
\newcommand\norm[1]{#1\!\!\downarrow\,} 
\newcommand\normE[2]{#1\!\!\downarrow_{#2}\ }
\newcommand\pub[1]{\mathsf{pub}(#1)}
\newcommand\sign[2]{\mathsf{sign}(#1,#2)}
\newcommand\blind[2]{\mathsf{blind}(#1,#2)}
\def\pubsym{\mathsf{pub}}
\def\signsym{\mathsf{sign}}
\def\blindsym{\mathsf{blind}}
\def\nameset{\mathsf{N}}
\def\varset{\mathsf{V}}
\def\conset{\Sigma_C}

\def\entail{\Vdash}
\def\seqsym{\vdash}
\def\decsym{{\mathrm{dec}}}

\def\Dscr{{\mathcal D}}

\def\Lscr{{\mathcal L}}
\def\Nscr{{\mathcal N}}
\def\Rscr{{\mathcal R}}
\def\Sscr{{\mathcal S}}

\def\doi{6 (3:12) 2010}
\lmcsheading%
{\doi}
{1--37}
{}
{}
{Oct.~30, 2009}
{Sep.~\phantom01, 2010}
{}   

\begin{document}

\title{A Proof Theoretic Analysis of Intruder Theories\rsuper*}
\author[A.~Tiu]{Alwen Tiu}
\author[R.~Gor\'e]{Rajeev Gor\'e}
\author[J.~Dawson]{Jeremy Dawson}
\address{
Logic and Computation Group \\ 
College of Engineering and Computer Science \\
The Australian National University 
}
\email{\{Alwen.Tiu, Rajeev.Gore, Jeremy.Dawson\}@rsise.anu.edu.au}
\keywords{AC convergent theories, sequent calculus,
  intruder deduction, security protocols}
\subjclass{F.3.1}
\titlecomment{{\lsuper*}An extended abstract has appeared
in the proceedings of the 2009 international
conference on Rewriting Techniques and Applications (RTA 2009)}

\begin{abstract}
  We consider the problem of intruder deduction in security protocol
  analysis: that is, deciding whether a given message $M$ can be
  deduced from a set of messages $\Gamma$ under the theory of blind
  signatures and arbitrary convergent equational theories modulo
  associativity and commutativity (AC) of certain binary
  operators. The traditional formulations of intruder deduction are
  usually given in natural-deduction-like systems and proving
  decidability requires significant effort in showing that the rules
  are ``local'' in some sense. By using the well-known translation
  between natural deduction and sequent calculus, we recast the
  intruder deduction problem as proof search in sequent calculus, in
  which locality is immediate. Using standard proof theoretic methods,
  such as permutability of rules and cut elimination, we show that the
  intruder deduction problem can be reduced, in polynomial time, to
  the elementary deduction problem, which amounts to solving certain
  equations in the underlying individual equational theories.  We show
  that this result extends to combinations of disjoint AC-convergent
  theories whereby the decidability of intruder deduction under the
  combined theory reduces to the decidability of elementary deduction
  in each constituent theory. Although various researchers have
  reported similar results for individual cases, our work shows that
  these results can be obtained using a systematic and uniform
  methodology based on the sequent calculus.  To further demonstrate
  the utility of the sequent-based approach, we show that, for
  Dolev-Yao intruders, our sequent-based techniques can be used to
  solve the more difficult problem of solving {\em deducibility
    constraints}, where the sequents to be deduced may contain gaps
  (or variables) representing possible messages the intruder may
  produce.  In particular, we show that there is a finite
  representation of all solutions to such a constraint problem.
\end{abstract}

\maketitle

\section{Introduction}

One of the fundamental aspects of the analysis of security protocols is the model
of the intruder that seeks to compromise the protocols. In many situations, 
such a model can be described in terms of a deduction system which gives a formal
account of the ability of the intruder to analyse and synthesize messages. 
As shown in many previous works 
(see, e.g., ~\cite{Amadio00CONCUR,Boreale01ICALP,Comon-Lundh03LICS,Chevalier03LICS}), 
finding attacks on protocols can often be framed as
the problem of deciding whether a certain formal expression is derivable in
the deduction system which models the intruder capability. The latter is sometimes called
the {\em intruder deduction problem}, or the (ground) reachability problem.  
A basic deductive account of the intruder's capability 
is based on the so-called Dolev-Yao model, which assumes perfect encryption. 
While this model has been applied fruitfully to many situations, 
a stronger model of intruders is needed to discover certain types of attacks.
For example, a recent survey \cite{Cortier06JCS} shows that attacks on several protocols 
used in real-world communication networks can be found by exploiting algebraic properties of 
encryption functions. 

The types of attacks mentioned in \cite{Cortier06JCS} have motivated
many recent works in studying models of intruders in which the
algebraic properties of the operators used in the protocols are taken
into account~\cite{Comon-Lundh03LICS,Chevalier03LICS,Abadi06TCS,Delaune06ICALP,Lafourcade07IC,Cortier07LPAR}.
In most of these, the intruder's capability is usually given as
a natural-deduction-like deductive system. As is common in natural
deduction, each constructor has a rule for introducing the constructor and
one for eliminating the constructor.
The elimination rule typically decomposes a term, reading
the rule top-down: {\em e.g.}, a typical elimination rule for a pair
$\spr M N$ of terms is:
$$
\infer[]
{\Gamma \vdash M}
{\Gamma \vdash \spr M N}
$$
Here, $\Gamma$ denotes a set of terms, which represents the terms
accumulated by the intruder over the course of its interaction with
participants in a protocol. While a natural deduction formulation of
deductive systems may seem ``natural'' and may reflect the meaning of
the (logical) operators, it does not immediately give us a proof
search strategy. Proof search means that we have to apply the rules
bottom up, and as the above elimination rule demonstrates, this
requires us to come up with a term $N$ which might seem arbitrary.
For a more complicated example, consider the following elimination
rule for {\em blind signatures} \cite{Fujioka92,KremerESOP05,Bernat06}.
$$
\infer[]
{\Gamma \vdash \sign M K}
{\Gamma \vdash \sign {\blind M R} K & \Gamma \vdash R}
$$
The basis for this rule is that the ``unblinding'' operation commutes
with signature.
Devising a proof search strategy in a natural deduction system
containing this type of rule does not seem trivial. 
In most of the works mentioned above, in order to show the decidability
results for the natural deduction system, one needs to prove that 
the system satisfies a notion of {\em locality}, 
i.e., in searching
for a proof for $\Gamma \vdash M$, one needs only to consider
expressions which are made of subterms from $\Gamma$ and
$M.$ 
In addition, one has to also deal with the complication that arises from
the use of the algebraic properties of certain operators.

In this work, we recast the intruder deduction problem as proof search
in sequent calculus. A sequent calculus formulation of Dolev-Yao
intruders was previously used by the first author in a formulation of
open bisimulation for the spi-calculus~\cite{Tiu07APLAS} to prove
certain results related to open bisimulation.
The current work takes this idea further to include richer theories.  Part
of our motivation is to apply standard techniques, which have been
well developed in the field of logic and proof theory, to the intruder
deduction problem. In proof theory, sequent calculus is commonly
considered a better calculus for studying proof search and
decidability of logical systems, in comparison to natural
deduction. This is partly due to the so-called ``subformula''
property (that is, the premise of every inference rule is made up of
subterms of the conclusion of the rule), which in most cases entails the decidability of the
deductive system. It is therefore rather curious 
that sequent calculus has not been more widely used in solving
intruder deduction. Some early work by Millen and Shmatikov, e.g., \cite{Millen01},
appears to incorporate aspects of sequent calculus inference rules
in their decision procedure for solving intruder deduction, but
apart from this work, we are not aware of any systematic
use of sequent calculus to solve the intruder deduction problem.
It is important to note that we do not think that sequent calculus is
a replacement for natural deduction as a {\em specification framework};
natural deduction is, naturally, a more intuitive framework to specify
an intruder's ability. What we propose here is an alternative way to
{\em structure proof search}, using known and widely used techniques
from proof theory.

We are mainly concerned with the ground intruder deduction problem (i.e., there are no
variables in terms) under the class of {\em AC-convergent
  theories}. These are equational theories that can be turned into
convergent rewrite systems, modulo associativity and commutativity of
certain binary operators. Many important theories for intruder
deduction fall into this category, e.g., theories for
exclusive-or~\cite{Comon-Lundh03LICS,Chevalier03LICS}, Abelian
groups~\cite{Comon-Lundh03LICS}, and more generally, certain classes
of monoidal theories~\cite{Cortier07LPAR}. We shall also present
a solution to the more difficult problem of {\em deducibility constraint
problems} (see Section~\ref{sec:constraint}), as a demonstration
of feasibility of the sequent-based techniques, but only for a 
restricted model of the intruder.

A summary of the main results we obtain: 
We show that the decidability of intruder deduction under AC-convergent theories can be
reduced, in polynomial time, 
to {\em elementary intruder deduction problems}, which involve
only the equational theories under consideration.
We show that the intruder deduction problem for a
combination of disjoint theories $E_1,\ldots,E_n$ can be reduced, in polynomial time, 
to the elementary deduction problem {\em for each theory $E_i$.} This
means that if the elementary deduction problem is decidable for each
$E_i$, then the intruder deduction problem under the combined theory
is also decidable.
We note that these decidability results are not really new,
although there are slight differences and improvements over the
existing works (see Section~\ref{sec:rel}).
Our contribution is more of a methodological nature. We
arrive at these results using rather standard proof theoretical
techniques, e.g., {\em cut-elimination} and permutability of inference
rules, in a uniform and systematic way. In particular, we obtain
locality of proof systems for intruder deduction, which is one of the
main ingredients to decidability results in 
\cite{Comon-Lundh03LICS,Chevalier03LICS,Delaune06ICALP,Delaune06IPL},
for a wide range of theories that cover those studied in these works. 
Note that these works deal with a more difficult problem of deducibility
constraints, which models {\em active intruders}.
We have not yet covered this more general problem for the intruder
models with AC convergent theories, although, as we mentioned above,
we do show a sequent-based solution to a restricted model of 
intruders (without AC theories). As future work, we plan to
extend our approach to deal with active intruders under richer intruder
models. 

The remainder of the paper is organised as follows. 
Section~\ref{sec:intruder} presents two systems for intruder theories,
one in natural deduction and the other in sequent calculus, and shows that the two systems are equivalent.
In Section~\ref{sec:cutelim}, the sequent system is shown to enjoy cut-elimination. 
In Section~\ref{sec:normal}, we show that cut-free sequent derivations can be transformed into a certain
normal form. Using this result, we obtain another ``linear'' sequent system, from which the polynomial
reducibility result follows. 
Section~\ref{sec:comb} shows that the sequent system 
in Section~\ref{sec:intruder} can be extended straightforwardly to cover any combination of 
disjoint AC-convergent theories,
and the same decidability results also hold for this extension. 
In Section~\ref{sec:constraint} we show that the sequent-based techniques, in particular
the normal form theorem, can be used to solve the more difficult problem of
solving deducibility constraints for Dolev-Yao intruders, which do not involve
any equational theories. 
The main results in Section~\ref{sec:constraint}, i.e., cut elimination and decision procedures
for both intruder deduction and deducibility constraints, have been formally verified
in Isabelle/HOL by the third author. 

This paper is a revised and extended version of a conference paper~\cite{Tiu09RTA}.
More specifically, we have added detailed proofs of the results stated in the
conference version and a new section on the sequent-based approach to solving 
deducibility constraint problems for Dolev-Yao intruders.

\section{Intruder deduction under AC-convergent theories}
\label{sec:intruder}

We consider in the following the problem of formalising, given a set of messages
$\Gamma$ and a message $M$, whether $M$ can be synthesized from the messages
in $\Gamma.$ We shall write this judgment as $\Gamma \seqsym M.$ This is sometimes called
the `ground reachability' problem or the `intruder deduction' problem in the literature. 

Messages are formed from names, variables and function symbols.  We
shall assume the following sets: a countably infinite set $\nameset$
of names ranged over by $a$, $b$, $c$, $d$, $m$ and $n$; a countably
infinite set $\varset$ of variables ranged over by $x$, $y$ and $z$;
and a finite set $\conset = \{\pubsym, \signsym, \blindsym,
\spr{\_}{\_}, \enc{\_}{\_}\}$ of symbols representing the {\em
  constructors}. Thus $ \pubsym$ is a public key constructor,
$\signsym$ is a constructor representing public key signature, $\blindsym$ is the
blinding encryption function (as in
\cite{Fujioka92,KremerESOP05,Bernat06}), $\spr{\_}{\_}$ is a pairing
constructor, and $\enc{\_}{\_}$ is the Dolev-Yao symmetric encryption
function. Note that the choice of the constructors here is not 
the most exhaustive one, in the sense that it does not cover
all commonly used Dolev-Yao types of constructors (e.g., hash, 
asymmetric encryption, etc.); we select a subset 
which we think is representative enough. Adding those extra
constructors to our model is straightforward, and the main results
of this paper should extend to these additions as well. 
Note also that for clarity of presentation, 
in presenting the deduction rules corresponding
to the encryption or signing operators, we do not attempt to
abstract them further, e.g., by presenting a generic form of
rules that could account for both encryption and signing (as they
both have a similar structure).  

In addition to constructors, we also assume a possibly empty equational
theory $E$, whose signature is denoted with $\Sigma_E.$ We require
that $\Sigma_C \cap \Sigma_E = \emptyset.$\footnote{This restriction
means that an intruder theory such as homomorphic encryption is excluded.
Nevertheless, it still covers a wide range of intruder theories.}
Function symbols (including
constructors) are ranged over by $f$, $g$ and $h$.  
The equational theory $E$ may contain any number of associative-commutative function
symbols, obeying the standard associative and commutative laws. 
However, for clarity of exposition, in this section, we shall restrict $E$ to contain at most
one associative-commutative symbol, which we denote with $\oplus$. 
Later in Section~\ref{sec:comb}, we shall consider the more general case where the equational
theory $E$ can contain an arbitrary number of AC symbols.
In any case, we restrict ourselves to equational
theories which can be represented by terminating and confluent rewrite
systems, modulo the associativity and commutativity of $\oplus.$ We
consider the set of messages generated by the following grammar
$$
\begin{array}{ll}
M, N := & a \mid x \mid \pub M \mid \sign M N \mid \blind M N 
\\      & \;\;\; \mid \spr M N \mid \enc M N \mid f(M_1,\ldots,M_k)
\end{array}
$$
where $f \in \Sigma_E.$
The operational meaning of each constructor will be defined by
their corresponding inference rules. Here we give an intuitive
explanation for each constructor. Note that the language of messages
as given above is untyped, but in the following explanation, it is
helpful to draw analogy from practices in security protocol analysis
to distinguish certain types of messages such as (public/private) keys, 
names, etc. 
The message $\pub M$ denotes the public key generated from a private key $M$; 
$\sign M N$ denotes a message $M$ signed with a private key $N$;
$\blind M N$ denotes a message $M$ encrypted with $N$ using a 
special blinding encryption; $\spr M N$ denotes a pair of messages;
and $\enc M N$ denotes a message $M$ encrypted with a key $N$ using
Dolev-Yao symmetric encryption. 
The blinding encryption has a special property that it commutes with
the $\signsym$ operation, i.e., one can ``unblind'' a signed blinded
message $\sign {\blind M r} k$ using the blinding key $r$  
to obtain $\sign M k.$ This aspect of the blinding encryption is 
reflected in its elimination rules, as we shall see later.
We denote with $V(M)$ the set of variables occurring in $M$. A message $M$ is {\em ground}
if $V(M) = \emptyset.$ 
In the following, we shall be mostly concerned with ground terms, so unless
stated otherwise, we assume implicitly that messages are ground.
The only exception is Proposition~\ref{prop:var-abs} and
Proposition~\ref{prop:var-abs2}
and Section~\ref{sec:constraint} where non-ground messages are also considered.

We shall use several notions of equality so we distinguish them using the following
notation: we shall write $M = N$ to denote syntactic equality, $M
\equiv N$ to denote equality modulo associativity and commutativity
(AC) of $\oplus$, and $M \approx_T N$ to denote equality modulo a
given equational theory $T$. We shall sometimes omit the subscript in
$\approx_T$ if it can be inferred from context.

\begin{rem}
Note that there is a choice on what function symbols one can regard
as constructors and what one can put into the equational theory.
At one extreme, we can consider all function symbols as part of the
equational theory, e.g., by introducing one or more ``destructor'' functions
for each constructor, and capture the intended meaning of each constructor
via equations. For example, for symmetric encryption, one could introduce
a decryption operator $\decsym$ satisfying:
$$
\dec {\enc M N} N \approx M,
$$
and for pairing, one could introduce the standard projection functions:
$$
\pi_1(\spr M N) \approx M
\qquad \mbox{ and } \qquad
\pi_2(\spr M N) \approx N.
$$
However, incorporating all function symbols into the equational theory in this manner
means that we lose the benefit of sequent calculus
in analysing the structures of deduction, as equational theories are less constrained than inference rules 
as far as proof search is concerned. 
Ideally, one would want to push {\em all} function symbols into the inference
system, but there appears to be no easy way to accomodate the associative-commutative
symbols. 
The set of constructors that we can accomodate in the inference system is obviously
larger than the one we consider here. Essentially, all equations that
involve constructor-destructor pairs 
that obey simple equations, like the ones for pairing above, can be
turned into appropriate introduction and elimination rules (in natural deduction)
for the constructors.
We leave as future work the exact characterisations of the equational theories that can be
absorbed into inference rules.
\end{rem}

Given an equational theory $E$, we denote with $R_E$ the set of
rewrite rules for $E$ (modulo AC). We write $M \to_{R_E} N$ when 
$M$ rewrites (modulo AC) to $N$ using one application of a rewrite rule in $R_E$. The definition
of rewriting modulo AC is standard and is omitted here (see, e.g., \cite{Baader98book} for
a definition). 
We recall one assumption about variables in rewrite rules that will be
used explicitly in some proofs in the following section: if $s \to_{R_E} t$
is a rewrite rule, then the variables in $t$ must occur in $s.$ 
The reflexive-transitive closure of $\to_{R_E}$  is denoted with $\rightarrow_{R_E}^*.$ 
We shall often remove the subscript $R_E$ when no confusion arises. 
A term $M$ is in {\em $E$-normal form} if $M \not \to_{R_E} N$ for any $N.$
We write $\normE M E$ to denote the normal form of $M$ with respect to 
the rewrite system $R_E$, modulo commutativity and associativity of $\oplus$.
Again, the index $E$ is often omitted when it is clear which equational theory we
refer to.
This notation extends straightforwardly to sets, e.g., $\norm \Gamma$ denotes the set obtained
by normalising all the elements of $\Gamma.$

A term $M$ is said to be {\em headed by } a symbol $f$ if $M = f(M_1,\ldots,M_k)$.
A term $M$ is an {\em $E$-alien term} if $M$ is headed by a symbol $f \not \in \Sigma_E.$
It is a {\em pure $E$-term} if it contains only symbols from $\Sigma_E$, names and variables.
A term $M$ is a {\em proper subterm} of $N$ if $M$ is a subterm of $N$ and $M \not = N.$
Given a term $M = f(M_1,\ldots, M_k)$, where $f$ is a constructor or a function symbol, 
the terms $M_1, \ldots, M_k$ are called the
{\em immediate subterms} of $M.$ 

An $E$-alien subterm $M$ of $N$ is said to be an {\em $E$-factor} of $N$ if 
there is another subterm $F$ of $N$ such that $M$ is an immediate
subterm of $F$ and $F$ is headed by a symbol $f \in \Sigma_E.$
This notion of a factor of a term is generalised to sets of terms
in the obvious way: a term $M$ is an  $E$-factor of $\Gamma$ if it is an $E$-factor of
a term in $\Gamma.$

\begin{exa}
The term $M = d \oplus (\spr c {\spr a b})$ has only one $E$-factor:
$\spr c {\spr a b}.$
Note that $\spr a b$ is not an $E$-factor of $M$, since no subterm
of $M$ containing $\spr a b$ as its immediate subterm is headed
by a symbol from $\Sigma_E.$ 
The subterm $d$ is not an $E$-factor of $M$ either, since it is not an $E$-alien term.
\end{exa}

A {\em context} is a term with holes.
We denote with $C^k[]$ a context with $k$-hole(s).
When the number $k$ is not important or can be inferred from context,
we shall write $C[\ldots]$ instead.
Viewing a context $C^k[]$ as a tree, each hole in the context
occupies a unique position among the leaves of the tree.
We say that a hole occurrence is the $i$-th hole of the context $C^k[]$
if it is the $i$-th hole encountered in
an inorder traversal of the tree representing $C^k[].$
An $E$-context is a context formed using only
the function symbols in $\Sigma_E.$ 
We write $C[M_1,\ldots,M_k]$ to denote the term resulting 
from replacing the holes in the $k$-hole context $C^k[]$ 
with $M_1, \ldots, M_k,$ where $M_i$ occupies the $i$-th hole in $C^k[].$ 

\paragraph{Natural deduction and sequent systems.}

The standard formulation of the judgment $\Gamma \seqsym M$ is usually given in
terms of a natural-deduction style inference system, as shown in
Figure~\ref{fig:nat}. We shall refer to this proof system as $\Nscr$
and write $\Gamma \entail_\Nscr M$ if $\Gamma \seqsym M$ is derivable in $\Nscr.$
The deduction rules for Dolev-Yao encryption are standard and can be found
in the literature, e.g., \cite{Boreale01ICALP,Comon-Lundh03LICS}. 
The blind signature rules are taken from the formulation given by Bernat and 
Comon-Lundh~\cite{Bernat06}. Note that the rule $\signsym_E$ assumes
implicitly that signing a message hides its contents. An alternative rule
without this assumption would be
$$
\infer[]
{\Gamma \seqsym M}
{\Gamma \seqsym \sign M K}
$$
The results of the paper also hold, with minor modifications, if we adopt this rule.

\begin{figure}[t]
\begin{center}
  \begin{tabular}[c]{l@{\quad}l@{\quad}l}
    $\infer[id]{\Gamma \seqsym M}
               {M \in \Gamma}$
  & $\infer[e_E]{\Gamma \seqsym M}
                {\Gamma \seqsym \enc M K & \Gamma \seqsym K}$
  & $\infer[e_I]{\Gamma \seqsym \enc M K}
                {\Gamma \seqsym M & \Gamma \seqsym K}$
\\[1em]
    $\infer[p_E]{\Gamma \seqsym M}
                {\Gamma \seqsym \spr M N}$
  & $\infer[p_E]{\Gamma \seqsym N}
                {\Gamma \seqsym \spr M N}$
  & $\infer[p_I]{\Gamma \seqsym \spr M N}
                {\Gamma \seqsym M & \Gamma \seqsym N}$
\\[1em]
  \multicolumn{2}{c}{
      $\infer[\signsym_E]{\Gamma \seqsym M}
                         {\Gamma \seqsym \sign M K & \Gamma \seqsym \pub K}$
    }
  &  $\infer[\signsym_I]{\Gamma \seqsym \sign M K}
                       {\Gamma \seqsym M & \Gamma \seqsym K}$
\\[1em]
    \multicolumn{2}{c}{
    $\infer[\blindsym_{E1}]{\Gamma \seqsym M}
                          {\Gamma \seqsym \blind M K & \Gamma \seqsym K}$
    }
  &
    $\infer[\blindsym_I]{\Gamma \seqsym \blind M K}
                        {\Gamma \seqsym M & \Gamma \seqsym K}$
\\[1em]
    \multicolumn{3}{c}{
     $\infer[\blindsym_{E2}]{\Gamma \seqsym \sign M K}
                       {\Gamma \seqsym \sign {\blind M R} K & \Gamma \seqsym R}$
    }
\\[1em]
    \multicolumn{2}{c}{
    $\infer[f_I, \hbox{ where } f \in \Sigma_E \qquad]
           {\Gamma \seqsym f(M_1,\ldots,M_n)}
           {\Gamma \seqsym M_1 & \cdots & \Gamma \seqsym M_n}$
    }
  & 
    \multicolumn{1}{c}{
    $\infer[\approx, \hbox{ where $M \approx_{E} N$}]
                          {\Gamma \seqsym M}
                          {\Gamma \seqsym N}$
    }
  \end{tabular}
\end{center}
\caption{System $\Nscr$: a natural deduction system for intruder deduction}
\label{fig:nat}
\end{figure}

\begin{figure}[t]
  \begin{center}
  \begin{tabular}[c]{ll}
   $\infer[id]{\Gamma \seqsym M}
              {
              \begin{array}{c}
               M \approx_{E} C[M_1,\ldots,M_k] \\
                \hbox{$C[\ ]$ an $E$-context, and $M_1,\ldots,M_k \in \Gamma$}
              \end{array}
              }$
  & $\infer[cut]{\Gamma \seqsym T}
                {\Gamma \seqsym M
                 & \Gamma, M \seqsym T}$
\\[1em]
  $\infer[p_L]{\Gamma, \spr M N \seqsym T}
              {\Gamma, \spr M N, M, N \seqsym T}$
  & $\infer[p_R]{\Gamma \seqsym \spr M N}
                {\Gamma \seqsym M & \Gamma \seqsym N}$
\\[1em]
  $\infer[e_L]{\Gamma, \enc M K \seqsym N}
              {\Gamma, \enc M K \seqsym K
                & \Gamma, \enc M K, M, K \seqsym N}$
  & $\infer[e_R]{\Gamma \seqsym \enc M K}
                {\Gamma \seqsym M & \Gamma \seqsym K}$
\\[1em]
  $\infer[\signsym_L, K \equiv L]{\Gamma, \sign M K, \pub L \seqsym N}
                                 {\Gamma, \sign M K, \pub L, M \seqsym N}$
  & $\infer[\signsym_R]{\Gamma \seqsym \sign M K}
                       {\Gamma \seqsym M & \Gamma \seqsym K}$
\\[1em]
  $\infer[\blindsym_{L1}]{\Gamma, \blind M K \seqsym N}
                        {\Gamma, \blind M K \seqsym K 
                         & \Gamma, \blind M K, M, K \seqsym N}$
  & $\infer[\blindsym_R]{\Gamma \seqsym \blind M K}
                       {\Gamma \seqsym M & \Gamma \seqsym K}$
\\[1em]
  \multicolumn{2}{c}{
    $\infer[\blindsym_{L2}]{\Gamma, \sign {\blind M R} K \seqsym N}
                          {\Gamma, \sign {\blind M R} K \seqsym R
                           &
                           \Gamma, \sign {\blind M R} K, \sign M K, R \seqsym N}$
  }
\\[1em]
  \multicolumn{2}{c}{
   $\infer[acut, \hbox{$A$ is an $E$-factor of $\Gamma \cup \{M\}$}]
           {\Gamma \seqsym M}
           {\Gamma \seqsym A & \Gamma, A \seqsym M}$
  }
\end{tabular}
\caption{System $\Sscr$: a sequent system for intruder deduction. }
\label{fig:msg}
  \end{center}
\end{figure}

A sequent $\Gamma \seqsym M$ is in {\em normal form} if $M$ and all the terms in $\Gamma$ are
in normal form. Unless stated otherwise, in the following we assume that sequents
are in normal form.
The sequent system for intruder deduction, under the equational theory $E$, 
is given in Figure~\ref{fig:msg}. We refer to this sequent system as $\Sscr$
and write $\Gamma \entail_\Sscr M$ to denote the fact that the sequent
$\Gamma \seqsym M$ is derivable in $\Sscr.$

Unlike natural deduction rules, sequent rules also allow introduction of terms on
the left hand side of the sequent. 
The rules $p_L,$ $e_L,$ $\signsym_L,$ $\blindsym_{L1},$ $\blindsym_{L2},$ and  $acut$ are called
{\em left introduction rules} (or simply {\em left rules}),
and the rules $p_R, e_R, \signsym_R, \blindsym_R$ are called {\em right introduction
rules} (or simply, {\em right rules}).
Notice that the rule $acut$ is very similar to $cut$, except that we have the
proviso that $A$ is an $E$-factor of the messages in the lower sequent. This is sometimes
called {\em analytic cut} in the proof theory literature. Analytic cuts are 
not problematic as far as proof search is concerned, since they still obey the
sub-formula property. 

We need the rule $acut$ because we do not have introduction rules for
function symbols in $\Sigma_E$, in contrast to natural deduction. This
rule is needed to ``abstract'' $E$-factors in a sequent (in the
sense of the variable abstraction technique common in unification theory, see
e.g., \cite{Schmidt-Schauss89,Baader96JSC}), which is needed to prove
that the cut rule is redundant.  For example, let $E$ be a theory containing 
only the associativity and the commutativity axioms for $\oplus$. 
Then the sequent ~ $ a, b \seqsym \spr a b \oplus a $ should be derivable without cut. 
Apart from the $acut$ rule, the only other way to derive this is by using the $id$
rule. However, $id$ is not applicable, since no $E$-context
$C[...]$ can obey $C[a,b] \approx \spr a b \oplus a$ 
because $E$-contexts can contain only symbols from
$\Sigma_E$ and thus cannot contain $\spr . .$. 
Therefore we need to abstract the term $\spr a b$ in the right 
hand side, via the $acut$ rule:
$$
\infer[acut]
{a, b \seqsym \spr a b \oplus a}
{
  \infer[p_R]
  {a, b \seqsym \spr a b}
  {\infer[id]{a,b\seqsym a}{} & \infer[id]{a,b \seqsym b}{}}
 & 
  \infer[id]
  {a, b, \spr a b \seqsym \spr a b \oplus a}
  {}
}
$$
The third $id$ rule instance (from the left) is valid because we have
$C[\spr a b, a] \equiv \spr a b \oplus a$, where
$C[.,.] = [.] \oplus [.].$

Derivability in the natural deduction system and in the sequent system
are related via the standard translation, i.e., right rules in sequent calculus
correspond to introduction rules in natural deduction and left rules correspond
to elimination rules. The straightforward 
translation from natural deduction to sequent calculus uses the cut rule.

\begin{rem}
Notice that the left rule for signing in the sequent calculus ($\signsym_L$)
and the left rule for symmetric encryption ($e_L$) have different forms,
although in the natural deduction system, their elimination rules are more
or less the same. We could indeed use the following alternative left-rule for
$\signsym_L$:
$$
\infer[\signsym_L']
{\Gamma, \sign M K, \seqsym N}
{\Gamma, \sign M K \seqsym \pub K & \Gamma, \sign M K, M, \pub K \seqsym N}
$$
It could be shown that $\signsym_L$ and $\signsym_L'$ are equivalent.
We prefer the former since it has a `nicer' form in that it satisfies
the subformula property. Notice also that in $\signsym_L$, we need
the proviso $K \equiv L$ because in the sequent rules, 
we do not quotient terms modulo AC. 
\end{rem}

In the following, given a derivation $\Pi$, we denote with $|\Pi|$ the {\em height} of $\Pi$,
i.e., the length of the longest branch in $\Pi$.

\begin{lem}[Weakening]
\label{lm:weak}
Let $\Pi$ be a derivation, in $\Sscr$, of $\Gamma \seqsym M$. 
If $\Gamma \subseteq \Gamma'$, then there exists an $\Sscr$-derivation
$\Pi'$ of $\Gamma' \seqsym M$ such that $|\Pi| = |\Pi'|$.
\end{lem}
\proof
By induction on $|\Pi|.$ \qed

\begin{lem}
\label{lm:nd-to-seq}
If the judgment $\Gamma \seqsym M$ is derivable in
the natural deduction system $\Nscr$ then
$\norm \Gamma \seqsym \norm M$ is derivable in the sequent system $\Sscr$.
\end{lem}
\proof
Let $\Pi$ be a natural deduction derivation of $\Gamma \seqsym M$.
We construct a sequent derivation
$\Pi'$ of $\norm \Gamma \seqsym \norm M$ by induction on $|\Pi|.$ 
The $id$ rule translates to the $id$ rule in sequent calculus;
the introduction rules
for constructors translate to the right-rules for the same constructors. 
If $\Pi$ ends with the $\approx$-rule,
then the premise and the conclusion of the rules translate to the same sequent,
hence $\Pi'$ is constructed by induction hypothesis. 
It remains to show the translations for 
the elimination rules and rules concerning $f \in \Sigma_E.$
\begin{enumerate}[$\bullet$]
\item Suppose $\Pi$ ends with $f_I$, for some $f \in \Sigma_E$:
$$
\infer[f_I]
{\Gamma \seqsym f(M_1,\ldots,M_k)}
{
 \deduce{\Gamma \seqsym M_1}{\Pi_1}
 & \cdots &
 \deduce{\Gamma \seqsym M_k}{\Pi_k}
}
$$
By induction hypothesis, we have sequent derivations $\Pi_i'$ of
$\norm \Gamma \seqsym \norm {M_i}$, for each $i \in \{1,\ldots,k\}$. 
Lemma~\ref{lm:weak}, applied to the $\Pi_i'$, gives us another sequent
derivation $\Pi_i''$ of $\norm \Gamma, \norm {M_1}, \ldots, \norm {M_{i-1}} 
  \seqsym \norm {M_i}$.
We note that the sequent 
$$
\norm \Gamma, \norm {M_1}, \ldots, \norm {M_k} \seqsym \norm{f(M_1,\ldots,M_k)}
$$
is derivable in the sequent system $\Sscr$ by an application of the $id$-rule 
since $C[] = f(\ldots)$ is an $E$-context.
The derivation $\Pi'$ is then constructed by successive applications of
the cut rule to this sequent with $\Pi_k'', \ldots, \Pi_1'',$ where the $i$-th
cut eliminates $\norm{M_i}$ from the conclusion by using the derivation
$\Pi_i''$ of $\norm \Gamma, \norm {M_1}, \ldots, \norm {M_{i-1}} 
  \seqsym \norm {M_i}$.

\item Suppose $\Pi$ ends with $p_E:$
$$
\infer[p_E]
{\Gamma \seqsym M}{\deduce{\Gamma \seqsym \spr M N}{\Pi_1}}
\qquad 
\infer[p_L]
{\norm \Gamma, \spr {\norm M}{\norm N} \seqsym \norm M}
{\infer[id]
{\norm \Gamma, \spr {\norm M}{\norm N}, \norm M, \norm N \seqsym \norm M}
{}}
$$
Note that $\norm {\spr M N} \equiv \spr {\norm M} {\norm N}$
and that the sequent $\norm \Gamma, \spr {\norm M}{\norm N} \seqsym \norm M$
is derivable in the sequent calculus $\Sscr$
(using an $id$ rule followed by a $p_L$-rule), as shown above right.
By the induction hypothesis,
we have a sequent derivation $\Pi_1'$ of
$\norm \Gamma \seqsym \spr {\norm M}{\norm N}$,
and so we can use the cut rule to get a sequent derivation of
$\norm \Gamma \seqsym \norm M.$

\item Suppose $\Pi$ ends with $e_E:$
$$
\infer[e_E]
{\Gamma \seqsym M}
{
  \deduce{\Gamma \seqsym \enc M N}{\Pi_1}
  &
  \deduce{\Gamma \seqsym N}{\Pi_2}
}
$$
By the induction hypothesis, we have a sequent derivation $\Pi_1'$
of $\norm \Gamma \seqsym \enc {\norm M} {\norm N}$ 
and a sequent derivation $\Pi_2'$ of $\norm \Gamma \seqsym \norm N.$
By Lemma~\ref{lm:weak}, we have a derivation $\Pi_3$ of
$\norm \Gamma, \enc {\norm M} {\norm N} \seqsym \norm N$,
where $|\Pi_3| = |\Pi_2'|$.
We construct a sequent derivation 
for the sequent
$$
\norm \Gamma, \enc {\norm M}{\norm N}, \norm N \seqsym \norm M
$$
by an application of $e_L$,
followed by two applications of $id$ (read upwards).
Then $\Pi'$ is constructed by applying the cut rule to this sequent
using $\Pi_3$ and $\Pi_1'$.

\item Suppose $\Pi$ ends with $\signsym_E$:
$$
\infer[\signsym_E]
{\Gamma \seqsym M}
{
 \deduce{\Gamma \seqsym \sign M K}{\Pi_1}
 &
 \deduce{\Gamma \seqsym \pub K}{\Pi_2}
}
$$
By induction hypothesis, we have a sequent derivation $\Pi_1'$
and a sequent derivation $\Pi_2'$ of, respectively,
$$
\norm \Gamma \seqsym \sign {\norm M} {\norm K} 
\qquad \hbox{ and } 
\qquad
\norm \Gamma \seqsym \pub {\norm K}.
$$
Let $\Pi_2''$ be a derivation of 
$$
\norm \Gamma, \sign {\norm M} {\norm K}  \seqsym \pub {\norm K}
$$
obtained by an application of Lemma~\ref{lm:weak} to $\Pi_2'.$
Let $\Pi_3$ be the derivation
$$
\infer[\signsym_L]
{\norm \Gamma, \sign {\norm M} {\norm K}, \pub {\norm K} \seqsym \norm M}
{
 \infer[id]
 {\norm \Gamma, \sign {\norm M} {\norm K}, \pub {\norm K}, \norm M \seqsym \norm M}
 {}
}
$$
Then $\Pi'$ is constructed by successive applications of 
cut with $\Pi_2''$ and cut with $\Pi_1'$ to $\Pi_3.$

\item The cases where $\Pi$ ends with $\blindsym_{E1}$ is
analogous to the case with $e_E$.

\item Suppose $\Pi$ ends with $\blindsym_{E2}$:
$$
\infer[\blindsym_{E2}]
{\Gamma \seqsym \sign M K}
{\deduce{\Gamma \seqsym \sign {\blind M R} K}{\Pi_1}
 & 
 \deduce{\Gamma \seqsym R}{\Pi_2}
}
$$
By induction hypothesis, we have a derivation $\Pi_1'$ and a derivation
$\Pi_2'$ of, respectively, 
$$
\norm \Gamma \seqsym \sign {\blind {\norm M} {\norm R}} {\norm K}
\qquad
\hbox{ and }
\qquad
\norm \Gamma \seqsym \norm R.
$$
Let $\Pi_3$ be the derivation
$$
\infer[\blindsym_{L2}]
{\norm \Gamma, \sign {\blind {\norm M} {\norm R}} {\norm K}  \seqsym \sign {\norm M} {\norm K}}
{
\deduce{\ldots \seqsym \norm R}{\Pi_2''}
&
\infer[id]
{\ldots, \sign {\norm M} {\norm K}, \norm R \seqsym \sign {\norm M} {\norm K}}
{}
}
$$
where $\Pi_2''$ is obtained from $\Pi_2'$ by weakening the sequent with
$$\sign {\blind {\norm M} {\norm R}} {\norm K}.$$
Then the derivation $\Pi'$ is constructed by a cut between $\Pi_1'$ and $\Pi_3.$
\qed
\end{enumerate}

\noindent For the case where the equational theory is empty, we
conjecture that the translation from natural deduction derivations to
sequent calculus derivations (with cuts) can be done in polynomial
time, as there are no duplication of derivation trees needed in the
translation. Note that in the translation, one needs to apply the
weakening lemma to weaken certain derivations, but this can be done in
linear time.  Note also that in the translation of elimination rules,
the cut rule is used to compose the inductively translated derivations
with new derivations. But the latter are all derivations of bounded
sizes (i.e., bounded by the size of the original sequent), hence they
can also be constructed in linear time, and the overall complexity
would still be bounded by polynomial time.

\begin{lem}
\label{lm:seq-to-nd}
If $\Gamma \seqsym M$, where $\Gamma \cup \{ M \}$ is a set of terms in normal
form, is derivable in the sequent system $\Sscr$ then
$\Gamma \seqsym M$ is derivable in the natural deduction system $\Nscr.$
\end{lem}
\proof
Let $\Pi$ be a sequent derivation of $\Gamma \seqsym M$. We construct a natural
deduction derivation $\Pi'$ of $\Gamma \seqsym M$ by induction on $\Pi.$
\begin{enumerate}[$\bullet$]
\item The right-introduction rules for $\Sscr$ map 
to the same introduction rules in $\Nscr.$
When $\Pi$ ends with such a rule, 
$\Pi'$ in this case is constructed straightforwardly from
the induction hypothesis using the introduction rules of $\Nscr.$
\item If $\Pi$ ends with an $id$ rule, i.e., $M \approx C[M_1,\ldots,M_k]$, for some
$M_1, \ldots, M_k \in \Gamma$ and $E$-context $C[..]$, we construct a derivation $\Pi_1$ of 
$\Gamma \seqsym C[M_1,\ldots,M_k]$ by induction on the context $C[\ldots]$. 
This is easily done using the $f_I$ introduction rule in $\Nscr.$
The derivation $\Pi'$ is then constructed from $\Pi_1$ by an application of the $\approx$-rule.

\item Suppose $\Gamma = \Gamma' \cup \{\spr U V\}$ and $\Pi$ ends with $p_L:$
$$
\infer[p_L]
{\Gamma', \spr U V \seqsym M}
{\deduce{\Gamma', \spr U V, U, V \seqsym M}{\Pi_1}}
$$
By induction hypothesis, we have an $\Nscr$-derivation $\Pi_1'$
of $\Gamma', \spr U V, U, V \seqsym M$.
We want an $\Nscr$-derivation $\Pi'$
of $\Gamma', \spr U V \seqsym M$ instead.
The $\Nscr$-derivation $\Pi'$ is constructed
inductively from $\Pi_1'$ by copying the same rule applications in $\Pi_1'$,
except when $\Pi_1'$ is either 
$$
\infer[id]
{\Gamma, U, V \seqsym U}
{}
~ \hbox{ or } ~
\infer[id]
{\Gamma, U, V \seqsym V}
{}
$$
in which case, $\Pi'$ is 
$$
\infer[p_E]
{\Gamma \seqsym U}
{
  \infer[id]
  {\Gamma \seqsym \spr U V}{}
}
~ \hbox{ and } ~
\infer[p_E]
{\Gamma \seqsym V}
{
  \infer[id]
  {\Gamma \seqsym \spr U V}{}
}
$$
respectively, since $\spr U V \in \Gamma$.

\item Suppose $\Gamma = \Gamma' \cup \{\enc U V\}$ and $\Pi$ ends with $e_L:$
$$
\infer[e_L]
{\Gamma', \enc U V \seqsym M}
{
 \deduce{\Gamma \seqsym V}{\Pi_1}
 &
 \deduce{\Gamma, U, V \seqsym M}{\Pi_2}
}
$$
By induction hypothesis, we have an $\Nscr$-derivation $\Pi_1'$ of 
$\Gamma \seqsym V$ and an $\Nscr$-derivation $\Pi_2'$ of
$\Gamma, U, V \seqsym M$.
The $\Nscr$-derivation $\Pi'$ of $\Gamma \seqsym M$
is then constructed inductively from $\Pi_2'$ 
by applying the same rules as in $\Pi_2'$, except when 
$\Pi_2'$ is either
$$
\infer[id]
{\Gamma, U, V \seqsym U}{}
~ \hbox{ or } ~
\infer[id]
{\Gamma, U, V \seqsym V}{}
$$
In the first case, $\Pi'$ is 
$$
\infer[e_E]
{\Gamma \seqsym U}
{
  \infer[id]{\Gamma \seqsym \enc U V}{}
  &
  \deduce{\Gamma \seqsym V}{\Pi_1'}
}
$$
and in the second case $\Pi'$ is simply $\Pi_1'$.

\item Suppose $\Gamma = \Gamma' \cup \{\sign N K, \pub L\}$ and 
$\Pi$ ends with $\signsym_L$:
$$
\infer[\signsym_L]
{\Gamma', \sign N K, \pub L \seqsym M}
{\deduce{\Gamma', \sign N K, \pub L, N \seqsym M}{\Pi_1}}
$$
where $L \equiv K$ (hence $L \approx K$).
By induction hypothesis, we have an $\Nscr$-derivation
$\Pi_1'$ of 
$$
\Gamma', \sign N K, \pub L, N \seqsym M.
$$
As in the previous case, the $\Nscr$-derivation $\Pi'$ of $\Gamma \seqsym M$
is constructed by imitating the rules of $\Pi_1'$, except for the following
$id$ case:
$$
\infer[id]
{\Gamma', \sign N K, \pub L, N \seqsym N}
{}
$$
which is replaced by 
$$
\infer[\signsym_E]
{\Gamma', \sign N K, \pub L \seqsym N}
{
 \infer[id]
 {\Gamma', \sign N K, \pub L \seqsym \sign N K}
 {}
 &
 \infer[\approx]
 {\Gamma', \sign N K, \pub L \seqsym \pub K}
 {
  \infer[id]
  {\Gamma', \sign N K, \pub L \seqsym \pub L}
  {}
 }
}
$$

\item The case where $\Pi$ ends with $\blindsym_{L1}$ is similar
to the case with $e_L.$

\item Suppose $\Gamma = \Gamma' \cup \{\sign {\blind N R} K  \}$
and  $\Pi$ ends with $\blindsym_{L2}$:
$$
\infer[\blindsym_{L2}]
{\Gamma', \sign {\blind N R} K \seqsym M}
{
\deduce{ \Gamma \seqsym R}{\Pi_1}
&
\deduce{ \Gamma, \sign N K, R \seqsym M}{\Pi_2}
}
$$
Similarly to the previous case, we apply the induction hypothesis to both
$\Pi_1$ and $\Pi_2$, obtaining $\Pi_1'$ and $\Pi_2'$. The derivation
$\Pi'$ is constructed by imitating the rules of $\Pi_2'$, but with the
following $id$ instances
$$
\infer[id]
{\Gamma', \sign N K, R \seqsym \sign N K}
{}
\qquad
\infer[id]
{\Gamma', \sign N K, R \seqsym R}{}
$$
replaced, respectively, by 
$$
\infer[]
{\Gamma \seqsym \sign N K}
{
 \infer[id]
 {\Gamma \seqsym \sign {\blind N R} K}{}
 &
 \infer[]
 {\Gamma \seqsym R}{\Pi_1'}
}
\quad
\hbox{ and }
\quad
\deduce{\Gamma \seqsym R}{\Pi_1'}.
$$

\item Suppose $\Pi$ ends with $acut$:
$$
\infer[acut]
{\Gamma \seqsym M}
{\deduce{\Gamma \seqsym A}{\Pi_1} 
 &
 \deduce{\Gamma, A \seqsym M}{\Pi_2}
}
$$
By induction hypothesis, we have an $\Nscr$-derivation
$\Pi_1'$ of $\Gamma \seqsym A$ and an $\Nscr$-derivation $\Pi_2'$
of $\Gamma, A \seqsym M.$ Again, as in the previous cases,
we construct $\Pi'$ inductively, on the height of $\Pi_2'$, 
by imitating the rules in $\Pi_2'$, except 
when $\Pi_2'$ ends with an instance of $id$ of the form
$$
\infer[id]
{\Gamma, A \seqsym A}
{}
$$
in which case, $\Pi'$ is $\Pi_1'.$

\item Suppose $\Pi$ ends with $cut$: this case is handled similarly
to the previous case. 
\qed
\end{enumerate}

\begin{prop}
\label{prop:S-equal-N}
The judgment $\Gamma \seqsym M$ is derivable in the natural deduction system
$\Nscr$ if and only if $\norm \Gamma \seqsym \norm M$
is derivable in the sequent system $\Sscr$.
\end{prop}
\proof
Immediate from Lemma~\ref{lm:nd-to-seq} and Lemma~\ref{lm:seq-to-nd}. \qed

\section{Cut elimination for \texorpdfstring{$\Sscr$}{S}}
\label{sec:cutelim}

We now show that the cut rule is redundant for $\Sscr$. 
\begin{defi}
An inference rule $R$ in a proof system $\Dscr$ is {\em admissible for $\Dscr$} if
for every sequent $\Gamma \vdash M$ derivable in $\Dscr$, there is a derivation
of the same sequent in $\Dscr$ without instances of $R$.
\end{defi}
The {\em cut-elimination} theorem for $\Sscr$ 
states that the cut rule is admissible for $\Sscr$. Before we proceed with the main cut elimination proof,
we first prove a basic property of equational theories and rewrite systems, which is concerned
with a technique called {\em variable abstraction} \cite{Schmidt-Schauss89,Baader96JSC}.

\subsection{Variable abstraction}

Given a normal term $M$, the {\em size} $|M|$ of $M$ is the number of function 
symbols, names and variables appearing in $M.$

In the following, we consider slightly more general equational
theories than in the previous section:  
each $AC$ theory $E$ can be a theory obtained from a disjoint combination of 
$AC$ theories $E_1, \ldots, E_k$, where each $E_i$ has at most
one AC operator $\oplus_i.$ This is so that we can reuse the results concerning
variable abstraction for a more general case later in Section~\ref{sec:comb}.

\begin{defi}
  Let $E$ be a disjoint combination of AC convergent theories
  $E_1,$ $\dots,$ $E_n$.  A term $M$ is a {\em quasi-$E_i$ term} if every
  $E_i$-alien subterm of $M$ is in $E$-normal form.
\end{defi}

\begin{exa}
Let $E = \{h(x,x) \approx x \}$.  
Then $h(\spr a b, c)$ is a quasi-$E$ term, whereas $h(\spr a b, \spr
{h(a,a)} b)$ is not, since its $E$-alien subterm $\spr {h(a,a)} b$ is not
in its $E$-normal form $\spr{a}{b}$.
Obviously, any $E$ normal term is a quasi-$E_i$ term.
\end{exa}

In the following, given an equational theory $E$, we assume the
existence of a function $v_E$, which assigns a variable from $\varset$
to each ground term such that $v_E(M) = v_E(N)$ if and only if $M
\approx_E N.$ In other words, $v_E$ assigns a unique variable to each
equivalence class of ground terms induced by $\approx_E.$

\begin{defi}
\label{def:var-abs}
Let $E$ be an equational theory obtained by disjoint combination of 
AC theories $E_1, \ldots, E_n$. 
The {\em $E_i$ abstraction function} $F_{E_i}$ is a function 
mapping ground terms to pure $E_i$ terms, defined recursively as follows:
$$
F_{E_i}(u) = 
\left\{
\begin{array}{ll}
u, & \hbox{ if $u$ is a name, }\\
f(F_{E_i}(u_1), \ldots, F_{E_i}(u_k)), & \hbox{if $u=f(u_1,\ldots,u_k)$ and $f \in \Sigma_{E_i}$,} \\
v_E(u), & \hbox{ otherwise.}
\end{array}
\right.
$$
\end{defi}

It can be easily shown that the function $F_{E_i}$ preserves the equivalence relation
$\equiv$. That is, if $M \equiv  N$ then $F_{E_i}(M) \equiv F_{E_i}(N)$.

\begin{lem}
\label{lm:var-abs}
Let $E$ be a disjoint combination of AC theories $E_1, \ldots, E_n$.
Let $M$ be a quasi-$E_i$ term. 
If $M \to_{R_E} N$ then $N$ is also a quasi-$E_i$ term
and $F_{E_i}(M) \to_{R_E} F_{E_i}(N).$
\end{lem}
\proof
By induction on the structure of $M$:
\begin{enumerate}[$\bullet$]
\item If $M$ is a name then the lemma holds vacuously.
\item Suppose $M = f(u_1,\ldots,u_k)$, where $f \in \Sigma_{E_i}.$
There are two cases to consider:
\begin{enumerate}[$-$]
\item The redex is in $u_j$. This case follows straightforwardly from the induction
hypothesis and the definition of $F_{E_i}$.
\item The redex is $M$. Then there must be a rewrite rule in $R_E$ of the form
$$
C[x_1,\ldots,x_n] \rightarrow C'[x_1,\ldots,x_n]
$$
where $C[..]$ and $C'[..]$ are $E_i$-context, such that
$$
M \equiv (C[x_1,\ldots,x_l])\sigma 
\qquad
\hbox{ and }
\qquad
N \equiv (C'[x_1,\ldots,x_l])\sigma
$$
for some substitution $\sigma.$
Note that since $M$ is a quasi-$E_i$ term, it follows that 
each $x_i\sigma$ is also a quasi-$E_i$ term. Hence $N$ must also be 
a quasi-$E_i$ term.
From the definition of $F_{E_i}$, we have the following
equality (we abbreviate $F_{E_i}$ as $F$):
$$
\begin{array}{ll}
F(M) & \equiv F(C[x_1,\ldots,x_l]\sigma) \\
& = C[F(x_1\sigma), \ldots, F(x_l\sigma)] \\
& = C[x_1,\ldots,x_l] \sigma'
\end{array}
$$
where $\sigma'$ is the substitution $\{F(x_1\sigma)/ x_1,\ldots, F(x_l\sigma)/x_l \}.$
Similarly, we can show that $F(N) \equiv C'[x_1,\ldots,x_l]\sigma'.$
Therefore, we have $F(M) \rightarrow_{R_E} F(N).$
\end{enumerate}

\item Suppose $M = g(u_1,\ldots,u_k)$ and $g\not \in \Sigma_{E_i}$.
Then $M$ is an $E_i$-alien subterm of $M$, and since $M$ is a quasi-$E_i$ term,
$M$ must be in $E$-normal form. Therefore no reduction is possible, hence the lemma
holds vacuously.\qed

\end{enumerate}

\begin{prop}
\label{prop:var-abs}
Let $E$ be a disjoint combination of $E_1,\ldots,E_n$. 
If $M$ is a quasi-$E_i$ term and $M \to^*_{R_E} N$, then
$N$ is a quasi-$E_i$ term and $F_{E_i}(M) \to^*_{R_E} F_{E_i}(N).$
\end{prop}
\proof This follows directly from Lemma~\ref{lm:var-abs}. \qed

\begin{prop}
\label{prop:var-abs2}
Let $E$ be a disjoint combination of $E_1,\ldots,E_n$. 
If $M$ and $N$ are quasi-$E_i$ terms and
$F_{E_i}(M) \to^*_{R_E} F_{E_i}(N)$, then 
$M \to^*_{R_E} N.$
\end{prop}
\proof
It is enough to show that this holds for the one-step rewrite
$F_{E_i}(M) \to_{R_E} F_{E_i}(N).$
This can be done by induction on the structure of $M$. 
In particular, we need to show that a rewrite rule that
applies to $F_{E_i}(M)$ also applies to $M$. 
Let $x_1,\ldots,x_k$ be the free variables in $F_{E_i}(M)$.
Let $M_1,\ldots,M_k$ be normal $E$-terms such that
$v_E(M_j) = x_j$ for each $j \in \{1,\ldots,k\}$, and
$$
\sigma = \{ M_1/x_1,\ldots,M_k/x_k\}.
$$
Then we can show by induction on the structure of $M$
and $N$, and using the fact that they are quasi-$E_i$ terms, 
that 
$$
F_{E_i}(M)\sigma \equiv M
\hbox{ and }
F_{E_i}(N)\sigma \equiv N.
$$
Note that for any rewrite rule in a rewrite system,
by definition, we have that all the variables free in
the right-hand side of the rule are also free in the
left-hand side.
Hence, the free variables of $F_{E_i}(N)$ are 
among the free variables in $F_{E_i}(M)$ since they are
related by rewriting.

Now suppose there is a rewrite rule in $R_E$
$$
C[y_1,\ldots,y_l] \rightarrow C'[y_1,\ldots,y_l]
$$
where $C[..]$ and $C'[..]$ are $E_i$-contexts, such
that $F_{E_i}(M) \equiv C[y_1,\ldots,y_l]\theta$
and $F_{E_i}(N) \equiv C'[y_1,\ldots,y_l]\theta$, for
some substitution $\theta.$
Then we have
$$
M \equiv F_{E_i}(M)\sigma \equiv (C[y_1,\ldots,y_l]\theta)\sigma \equiv C[y_1,\ldots,y_l](\theta \circ \sigma)
$$
and
$$
N \equiv F_{E_i}(N)\sigma \equiv (C'[y_1,\ldots,y_l]\theta)\sigma \equiv C'[y_1,\ldots,y_l](\theta \circ \sigma).
$$
Hence we also have $M \to_{R_E} N.$
\qed

\subsection{Cut elimination}

We now show some important proof transformations needed to prove cut elimination,
i.e., in an inductive argument to reduce the size of cut terms. 
In the following, when we write that a sequent $\Gamma \vdash M$ is derivable, 
we mean that it is derivable in
the proof system $\Sscr$, with a fixed AC theory $E$. Note that
here the equational theory $E$ contains at most one AC symbol.

\begin{lem}
\label{lm:equiv}
Let $\Pi$ be a derivation of $M_1, \ldots, M_k \vdash N.$ Then for any
$M_1',$ $\ldots,$ $M_k'$ and $N'$
such that $M_i \equiv M_i'$ and $N \equiv N'$, there is a derivation $\Pi'$
of $M_1', \ldots, M_k' \vdash N'$ such that $|\Pi| = |\Pi'|.$
\end{lem}
\proof
By induction on $|\Pi|.$ \qed

\begin{lem}
\label{lm:decomp1}
Let $X$ and $Y$ be terms in normal form and let $f$ be a binary
constructor.  If $\Gamma, f(X,Y) \vdash M$ is cut-free derivable, then
$\Gamma, X, Y \vdash M$ has a cut-free derivation.
\end{lem}
\proof
Let $\Pi$ be a cut-free derivation of $\Gamma, f(X,Y) \vdash M$. We construct a cut-free derivation
$\Pi'$ of $\Gamma, X, Y \vdash M$ by induction on $|f(X,Y)|$ with subinduction on $|\Pi|.$ 
The only non-trivial cases are when $\Pi$ ends with $\blindsym_{L2}$, acting on $f(X,Y)$,
and when $\Pi$ ends with $id$ and $f(X,Y)$ is used in the rule.
We examine these cases in more detail below.

\begin{enumerate}[$\bullet$]
\item Suppose $\Pi$ ends with $\blindsym_{L2}$, acting on $f(X,Y)$, i.e.,
$f = \signsym$ and $X = \blind N R$:
$$
\infer[\blindsym_{L2}]
{\Gamma, \sign {\blind N R} Y \vdash M}
{
\deduce{\Gamma, \sign {\blind N R} Y \vdash R}{\Pi_1}
&
\deduce{\Gamma, \sign {\blind N R} Y, \sign N Y, R \vdash M}{\Pi_2}
}
$$
Applying the inner induction hypothesis on derivation height 
to $\Pi_1$ and $\Pi_2$ we obtain two derivations $\Pi_1'$ and $\Pi_2'$ of
$$
\deduce{\Gamma, \blind N R, Y \vdash R}{\Pi_1'}
\quad \hbox{ and }
\quad
\deduce{\Gamma, \blind N R, Y, \sign N Y, R \vdash M}{\Pi_2'}
$$ 
Next we apply the outer induction hypothesis on the size of $f(X,Y)$ to 
decompose $\sign N Y$ in the latter sequent to get a derivation $\Pi_2''$ of
$$
\deduce{\Gamma, \blind N R, N, Y, R \vdash M}{\Pi_2''}
$$
The derivation $\Pi'$ is constructed as follows:
$$
\infer[\blindsym_{L1}]
{\Gamma, \blind N R, Y \vdash M}
{
\deduce{\Gamma, \blind N R, Y \vdash R}{\Pi_1'}
&
\deduce{\Gamma, \blind N R, N, Y, R \vdash M}{\Pi_2''}
}
$$
\item Suppose $\Pi$ ends with $id$. The only non-trivial case is 
when $f(X,Y)$ is active in the rule, that is, we have
$$
M \approx C[f(X,Y)^n, M_1, \ldots, M_k]
$$
where $M_1, \ldots, M_k \in \Gamma$, $C[\ldots]$ is an $E$-context and 
$f(X,Y)$ fills $n$-holes in $C[\ldots].$ We distinguish several cases:
\begin{enumerate}[$-$]
\item There is an $E$-factor $A$ of $M \cup \Gamma$ such that $f(X,Y) \equiv A.$
Note that in this case $A$ must be of the form $f(X',Y')$ for some $X' \equiv X$
and $Y' \equiv Y.$
In this case, $\Pi'$ is constructed as follows:
$$
\infer[acut]
{\Gamma, X,Y \vdash M}
{
  \deduce{\Gamma, X,Y \vdash f(X',Y')}{\Xi}
  &
  \infer[id]
  {\Gamma, X,Y, f(X',Y') \vdash M}{}
}
$$
where $\Xi$ is a derivation formed using $id$ and the right rules for the constructor
$f$.

\item Suppose that there is no $E$-factor $A$ of $M \cup \Gamma$ such that $A \equiv f(X,Y).$ 
Note that since $M$ is in normal form, we have
$$
C[f(X,Y)^n,M_1,\ldots,M_k] \to^* M
$$
and both $C[f(X,Y)^n,M_1,\ldots,M_k]$ and $M$ are quasi-$E$ terms. 

Let $x = v(f(X,Y))$. 
It follows from Proposition~\ref{prop:var-abs} that
$$
F_E(C[f(X,Y)^n, M_1,\ldots, M_k]) = C[x^n, F_E(M_1), \ldots, F_E(M_k)] \to^* F_E(M).
$$
Since no factors of $M$ and $M_1,\ldots,M_k$
are equivalent to $f(X,Y)$, $x$ obviously does not appear in
any of $F_E(M)$, $F_E(M_1), \ldots, F_E(M_k)$.
Now let $a$ be a name that does not occur in $\Gamma$, $X$, $Y$ or $M$.
Since rewriting is invariant under variable/name substitution, by substituting
$a$ for $x$ in the above sequence of rewrites, we have
$$
F_E(C[a^n, M_1,\ldots, M_k]) = C[a^n, F_E(M_1), \ldots, F_E(M_k)] \to^* F_E(M).
$$
Now by Proposition~\ref{prop:var-abs2}, we have
$$
C[a^n, M_1, \ldots, M_k] \to^* M.
$$
By substituting $X$ for $a$ in this sequence, we have
$$
C[X^n,M_1,\ldots,M_k] \longrightarrow_{\Rscr}^* M. 
$$
Thus, in this case, $\Pi'$ is constructed by an application of $id.$
\qed
\end{enumerate}
\end{enumerate}

\begin{lem}
\label{lm:decomp2}
Let $X_1,\ldots, X_k$ be terms in normal form and let 
$\Pi$ be a cut-free derivation of $\Gamma, \norm{f(X_1,\ldots,X_k)} \vdash M$, where $f \in \Sigma_E.$
Then there exists a cut-free derivation $\Pi'$ of $\Gamma, X_1,\ldots,X_k \vdash M.$ 
\end{lem}
\proof
By induction on $|\Pi|.$ The cases where $\Pi$ ends with $id$, or rules in which
$\norm{f(X_1,\ldots,X_k)}$ is not principal, are trivial. The other cases, where
$\Pi$ ends with a rule applied to the term $\norm{f(X_1,\ldots,X_k)},$ are given in the following.
\begin{enumerate}[$\bullet$]
\item Suppose $\Pi$ ends with $p_L$ on $\norm{f(X_1,\ldots,X_k)}.$ This means
that $\norm{f(X_1,\ldots,X_k)}$ is a pair $\spr U V$ for
some $U$ and $V$, and $\Pi$ is 
$$
\infer[p_L]
{\Gamma, \spr U V \seqsym M}
{
 \deduce{\Gamma, \spr U V, U, V \seqsym M}{\Xi}
}
$$
We have that
$$
f(X_1,\ldots,X_k) \to^* \spr U V.
$$
Let $x = F_E(\spr U V)$. By Proposition~\ref{prop:var-abs}, we have
$$
f(F_E(X_1,),\ldots, F_E(X_k)) \to^* x.
$$
Obviously, $x$ has to occur in $F_E(X_i)$ for some $X_i$. Without loss
of generality, assume that $i=1.$ 
This means that there exists an $E$-alien subterm $A$ of $X_1$ 
such that $A = \spr {U'} {V'}$ and
$U \equiv U'$ and $V \equiv V'$. There are two cases to consider.
\begin{enumerate}[$-$]
\item $A$ is a factor of $X_1$. 
Then $\Pi'$ is the derivation: 
$$
\infer[acut]
{\Gamma, X_1, \ldots, X_k \seqsym M}
{
  \infer[id]
  {\Gamma, X_1,\ldots, X_k \seqsym \spr {U'} {V'}}{}
  &
  \deduce{\Gamma, X_1,\ldots,X_k, \spr {U'} {V'} \seqsym M}{\Pi_1}
}
$$
The instance of $id$ above is valid since 
$\spr {U'} {V'} \equiv \spr U V \approx f(X_1,\ldots,X_k).$
The derivation $\Pi_1$ is obtained by weakening 
$\Pi$ with $X_1,\ldots, X_k$  and applying Lemma~\ref{lm:equiv} to replace
$\spr U V$ with its equivalent $\spr {U'} {V'}$.%

\item $A$ is not a factor of $X_1$. This can only mean that 
either $X_1 = A$ or that every occurrence of $A$ in $X_1$ is 
as immediate subterm of another $E$-alien subterm. 
The latter would mean that $A$ would not be abstracted by $F_{E_i}$ at all, 
contradicting the assumption that it is. So it must be the case
that $X_1 = A.$
Then $\Pi'$ is the derivation
$$
\infer[p_L]
{\Gamma, \spr{U'}{V'}, X_2,\ldots, X_k \seqsym M}
{
 \deduce{\Gamma, \spr{U'}{V'}, U',V', X_2,\ldots, X_k \seqsym M}{\Xi'}
}
$$
where $\Xi'$ is obtained by weakening $\Xi$ with $X_2,\ldots,X_k$,
and then applying Lemma~\ref{lm:equiv} to replace
$U$ and $V$ with their equivalent $U'$ and $V'$.
\end{enumerate}\medskip
The cases where $\norm{f(X_1,\ldots,X_k)}$ is headed with some other constructor
are proved analogously.

\item Suppose $\Pi$ ends with $acut$ which abstracts an $E$-factor of $\norm{f(X_1,\ldots,X_k)}$:
$$
\infer[acut]
{\Gamma, \norm{f(X_1,\ldots,X_k)} \vdash M}
{
\deduce{\Gamma' \vdash A}{\Pi_1}
&
\deduce{\Gamma', A \vdash M}{\Pi_2}
}
$$
where $A$ is an $E$-factor of $\norm{f(X_1,\ldots,X_k)}$
and $\Gamma' = \Gamma \cup \{\norm{f(X_1,\ldots,X_k)}\}$. 
In this case, we have that
$$
\norm{f(X_1,\ldots,X_k)} = C[g(\dots,A,\dots)]
$$
for some context $C[]$ and some $g \in \Sigma_E.$
By Proposition~\ref{prop:var-abs}, we have
$$
f(F_E(X_1,),\ldots, F_E(X_k)) \to^* F_E(C[g(\dots,A,\dots)]).
$$
We have a couple of cases to analyse, depending on whether
that particular occurrence of $A$ is abstracted by $F_E$
or not (i.e., if $g(\dots,A,\dots)$ is nested inside another
$E$-alien subterm). In both cases, it can be shown that
there exist $A' \equiv A$ and some $X_i$ such that either 
$A' = X_i$ or $A'$ is an $E$-factor of $X_i.$
For the latter case, $\Pi'$ is constructed as follows:
$$
\infer[acut]
{\Gamma, X_1, \ldots, X_k \vdash M}
{
  \deduce{\Gamma'' \vdash A'}{\Pi_1'}
  &
  \deduce{\Gamma'',A' \vdash M}{\Pi_2'}
}
$$
where $\Gamma'' = \Gamma \cup \{X_1,\ldots,X_k\}$ and $\Pi_1'$ and $\Pi_2'$ are obtained
by applying the induction hypothesis on $\Pi_1$ and $\Pi_2$, followed by
applications of Lemma~\ref{lm:equiv} to replace $A$ with its equivalent $A'.$
If $X_1 = A'$ then $\Pi'$ is obtained by weakening $\Pi_2$ with $X_2,\ldots,X_k$,
followed by an application of Lemma~\ref{lm:equiv} to replace $A$ with $A'.$
\qed
\end{enumerate}

\begin{lem}
\label{lm:decomp3}
Let $M_1, \ldots, M_k$ be terms in normal form and let $C[\ldots]$ be
a $k$-hole $E$-context. If $\Gamma, \norm{C[M_1,\ldots,M_k]} \vdash M$
is cut-free derivable, then so is $\Gamma, M_1, \ldots, M_k \vdash M$.
\end{lem}
\proof
By induction on the size of $C[\ldots]$, Lemma~\ref{lm:equiv} and Lemma~\ref{lm:decomp2}. \qed

One peculiar aspect of the sequent system $\Sscr$ is that in the
introduction rules for encryption functions (including blind
signatures), there is no switch of polarities for the encryption
key. For example, in the introduction rules for $\enc M K$, on both the
left and the right, the key $K$ appears on the right hand
side of a premise of the rule. This means that there is no exchange
of information between the left and the right hand side of sequents,
unlike typical implication rules in logic. This gives rise to an
easy cut elimination proof, where we need only to measure the
complexity of the left premise of a cut in determining the cut
rank.

\begin{thm}
\label{thm cut elim}
The cut rule is admissible for $\Sscr$.
\end{thm}
\proof
We give a set of transformation rules for derivations ending with cuts and show
that given any derivation, there is a sequence of reductions that applies to
this derivation, and terminates with a cut free derivation with the same end sequent. 
This is proved by induction on the height of the {\em left premise derivation} immediately above the cut rule. 
This measure is called the {\em cut rank}. As usual in cut elimination, we proceed by eliminating
the topmost instances of cut with the highest rank. So in the following, we suppose a given
derivation $\Pi$ ending with a cut rule, which is the only cut in $\Pi$, and then
show how to transform this to a cut free derivation $\Pi'.$

The cut reduction is driven by the left premise derivation of the cut. 
We distinguish several cases, based on the last rule of the left premise derivation. 
\begin{enumerate}[(1)]
\item Suppose the left premise of $\Pi$ ends with either $p_R$, $e_R$, $\signsym_R$
or $\blindsym_R$,  thus $\Pi$ is 
$$
\infer[cut]
{\Gamma \vdash R}
{
 \infer[\rho]
 {\Gamma \vdash f(M, N)}
 {\deduce{\Gamma \vdash M}{\Pi_1} & \deduce{\Gamma \vdash N}{\Pi_2}}
 &
 \deduce{\Gamma, f(M, N) \vdash R}{\Pi_3}
}
$$
where $f$ is a constructor and $\rho$ is its right introduction rule.
By Lemma~\ref{lm:decomp1}, 
we have a cut free derivation $\Pi_3'$ of 
$\Gamma, M, N \vdash R.$ By applying Lemma~\ref{lm:weak} to $\Pi_2$, we also have a cut-free derivation
$\Pi_2'$ of $\Gamma, M \vdash N$ such that $|\Pi_2| = |\Pi_2'|.$
The above cut is then reduced to
$$
\infer[cut]
{\Gamma \vdash R}
{
 \deduce{\Gamma \vdash M}{\Pi_1}
 &
 \infer[cut]
 {\Gamma, M \vdash R}
 {
  \deduce{\Gamma, M \vdash N} {\Pi_2'}
  &
  \deduce{\Gamma, M, N \vdash R}{\Pi_3'}
 }
}
\enspace .
$$
These two cuts can then be eliminated by induction hypothesis since their
left premises are of smaller height than the left premise of $\Pi.$

\item Suppose the left premise of the cut ends with a left rule 
acting on $\Gamma.$ 
We show here the case where the left-rule has only one premise; generalisation 
to the other case (with two premises) is straightforward. Therefore $\Pi$ is of the form:
$$
\infer[cut]
{\Gamma \vdash R}
{
 \infer[\rho]
 {\Gamma \vdash M}
 {
  \deduce{\Gamma' \vdash M}{\Pi_1}
 }
 &
 \deduce{\Gamma, M \vdash R}{\Pi_2}
}
$$
By inspection of the inference rules in Figure~\ref{fig:msg}, it is clear that 
in the rule $\rho$ above, we have $\Gamma \subseteq \Gamma'$. We can therefore weaken $\Pi_2$
to a derivation $\Pi_2'$ of $\Gamma', M \vdash R$ with $|\Pi_2| = |\Pi_2'|$. 
The cut is then reduced as follows.
$$
\infer[\rho]
{\Gamma \vdash R}
{
 \infer[cut]
 {\Gamma' \vdash R}
 {
   \deduce{\Gamma' \vdash M}{\Pi_1}
   &
   \deduce{\Gamma', M \vdash R}{\Pi_2}
 }
}
$$
The cut rule above $\rho$ can be eliminated by induction hypothesis, 
the height of the left premise of the cut is smaller than that of the left
premise of the original cut. 

\item Suppose the left premise of the cut ends with $acut$, but using an $E$-factor
of the right hand side of the sequent, i.e., $\Pi$ is
$$
\infer[cut]
{\Gamma \vdash R}
{
  \infer[acut]
  {\Gamma \vdash C[A]}
  {
    \deduce{\Gamma \vdash A}{\Pi_1}
    &
    \deduce{\Gamma, A \vdash C[A]}{\Pi_2}
  }
  &
  \deduce{\Gamma, C[A] \vdash R}{\Pi_3}
}
$$
Then this derivation reduces to: 
$$
\infer[cut]
{\Gamma \vdash R}
{
  \deduce{\Gamma \vdash A}{\Pi_1}
  &
  \infer[cut]
  {\Gamma, A \vdash R}
  {
    \deduce{\Gamma, A \vdash C[A]}{\Pi_2}
    &
    \deduce{\Gamma, A, C[A] \vdash R}{\Pi_3'}
  } 
}
$$ 
The derivation $\Pi_3'$ is obtained by weakening $\Pi_3$ with $A$ (Lemma~\ref{lm:weak}).
Both cuts can be removed by induction hypothesis (the upper cut followed by the lower cut).

\item Suppose the left premise of the cut ends with the $id$-rule:
$$
\infer[cut]
{\Gamma \vdash R}
{
 \infer[id]
 {\Gamma \vdash M}
 {}
 &
 \deduce{\Gamma, M \vdash R}{\Pi_1}
}
$$
where $M = \norm{C[M_1,\ldots,M_k]}$ and 
$M_1, \ldots, M_k \in \Gamma.$ 
In this case, we apply Lemma~\ref{lm:decomp3} to $\Pi_1$, hence we get
a cut free derivation $\Pi'$ of 
$\Gamma \vdash R.$
\qed
\end{enumerate}

\section{Normal derivations and decidability}
\label{sec:normal}

We now turn to the question of the decidability of the deduction problem
$\Gamma \vdash M.$ This problem is known to be decidable for several AC theories, 
e.g., exclusive-or, abelian groups and their extensions with a homomorphism
axiom~\cite{Comon-Lundh03LICS,Chevalier03LICS,Delaune06ICALP,Delaune06IPL,Abadi06TCS}. 
What we would like to show here is how the decidability
result can be reduced to a more elementary decision 
problem, defined as follows.

\begin{defi}
Given an equational theory $E$, the {\em elementary deduction problem} for $E$, written $\Gamma \Vdash_E M$, 
is the problem of deciding whether the $id$ rule is applicable
to the sequent $\Gamma \vdash M$ (by checking 
whether there exists an $E$-context
$C[\ldots]$ and terms $M_1,\ldots, M_k \in \Gamma$ such that
$C[M_1,\ldots,M_k] \approx_{E} M$).
\end{defi}

Note that as a consequence of Proposition~\ref{prop:var-abs} and Proposition~\ref{prop:var-abs2},
in checking elementary deducibility, it is enough to consider the pure $E$ equational
problem where all $E$-alien subterms are abstracted, i.e., we have
$$
C[M_1, \ldots, M_k] \approx_{E} M
\qquad
\mbox{iff}
\qquad
C[F_E(M_1), \ldots, F_E(M_k)] \approx_E F_E(M).
$$
Our notion of elementary deduction corresponds roughly to the notion of ``recipe''
in \cite{Abadi06TCS}, but we note that the notion of a recipe is a stronger one, since it
bounds the size of the equational context.

The cut free sequent system does not strictly speaking enjoy the ``sub-formula'' property,
i.e., in $\blindsym_{L2}$, the premise sequent has a term which is not a subterm of
any term in the lower sequent. However, it is easy to see that, reading the rules bottom up,
we only ever introduce terms which are smaller than the terms in the lower sequent.
Thus a naive proof search strategy which non-deterministically tries all applicable rules
and avoids repeated sequents will eventually terminate. 
This procedure is of course rather expensive. 
We show that we can obtain a better complexity result by analysing the structure of cut-free derivations. 
Recall that the rules $p_L, e_L, \signsym_L, \blindsym_{L1}, \blindsym_{L2}$ and $acut$
are called left rules (the other rules are right rules). 
Central to the decidability results in this section is the notion of a normal
derivation, given in the following definition.

\begin{defi}
A cut-free derivation $\Pi$ is said to be a {\em normal derivation} if 
it satisfies the following conditions: 
\begin{enumerate}[(1)]
\item no left rule appears above a right rule;
\item no left rule appears immediately above the left-premise of a
  branching left rule (i.e., all left rules except $p_L$ and
  $\signsym_L$).
\end{enumerate}
\end{defi}

\begin{lem}
\label{lm:perm1}
Let $\Pi$ be a cut-free derivation of $\Gamma \vdash M.$ Then there is a cut-free 
derivation of the same sequent such that all the right rules
appear above left rules.
\end{lem}
\proof
We permute any offending right rules up over any left
rules. This is done by induction on the number of occurrences of the offending rules.
We first show the case where $\Pi$ has at most one offending right rule.
In this case, we show, by induction on the height of $\Pi$, that any offending
right-introduction rule can be permuted up in the derivation tree until it 
is above any left-introduction rule. 
We show here a non-trivial case involving $acut$; the others are treated analogously.
Suppose $\Pi$ is as shown below at left
where $\rho$ denotes a right introduction rule for the constructor $f$
and $A$ is an $E$-factor of $\Gamma \cup \{ M\}$.
By the weakening lemma (Lemma~\ref{lm:weak}), we have a derivation
$\Pi_3'$ of $\Gamma, A \vdash N$ with $|\Pi_3'| = |\Pi_3|.$ The
original derivation $\Pi$ is
then transformed into the derivation shown below at right:
$$
\infer[\rho]
{\Gamma \vdash f(M,N)}
{
 \infer[acut]
 {\Gamma \vdash M}
 {
  \deduce{\Gamma \vdash A}{\Pi_1}
  &
  \deduce{\Gamma, A \vdash M}{\Pi_2}
 }
 &
 \deduce{\Gamma \vdash N}{\Pi_3}
}
\qquad
\infer[acut]
{\Gamma \vdash f(M,N)}
{
 \deduce{\Gamma \vdash A}{\Pi_1}
 &
 \infer[\rho]
 {\Gamma, A \vdash f(M,N)}
 {
  \deduce{\Gamma,A \vdash M}{\Pi_2}
  &
  \deduce{\Gamma, A \vdash N}{\Pi_3'}
 }
}
$$ 

The rule $\rho$ in the right premise can then be further permuted up (i.e., if
$\Pi_2$ or $\Pi_3'$ ends with a left rule) by induction hypothesis. 

The derivation $\Pi'$ is then constructed by repeatedly applying the above
transformation to the topmost offending rules until all of them appear
above left-introduction rules.
\qed

\begin{prop}
\label{prop:norm}
If $\Gamma \vdash M$ is derivable then it has a normal derivation. 
\end{prop}
\proof
Let $\Pi$ be a cut-free derivation of $\Gamma \vdash M$.
By Lemma~\ref{lm:perm1},
we can assume without loss of generality that all the right rules in
$\Pi$ appear above the left rules. 
We construct a normal derivation $\Pi'$ of the same sequent by induction
on the number of offending left rules in $\Pi$.

We first consider the case where $\Pi$ has at most one offending left rule. 
Let $\Xi$ be a subtree of $\Pi$ where the offending rule occurs, i.e., 
$\Xi$ ends with a branching left rule, whose left premise derivation ends with
a left rule. We show by induction on the height of the left premise derivation of 
the last rule in $\Xi$ that $\Xi$ can be transformed into a normal derivation.
There are two cases to consider: one in which the left premise derivation ends with
a branching left rule and the other where it ends with a non-branching left rule.
We consider the former case here, the latter can be dealt with analogously.
So suppose $\Xi$ is of the form:
$$
\infer[L_1]
{\Gamma_1 \vdash M'}
{
  \infer[L_2]
  {\Gamma_1 \vdash N_1}
  {
    \deduce{\Gamma_1 \vdash N_2}{\Pi_1}
    &
    \deduce{\Gamma_2 \vdash N_1}{\Pi_2} 
  }
  &
  \deduce{\Gamma_3 \vdash M'}{\Pi_3}
}
$$
where $L_1$ is a left rule, and $\Pi_1$, $\Pi_2$ and $\Pi_3$ are normal derivations, 
$\Gamma_2 \supseteq \Gamma_1$ and $\Gamma_3 \supseteq \Gamma_1.$ We first weaken $\Pi_3$ into a derivation 
$\Pi_3'$ of $\Gamma_4 \vdash M'$, where $\Gamma_4 = \Gamma_2 \cup \Gamma_3$. 
Such a weakening can be easily shown to not 
affect the shape of the derivations (i.e., it does not introduce or
remove any rules in $\Pi_3$). $\Xi$ is then transformed into 
$$
\infer[L_2]
{\Gamma_1 \vdash M'}
{
  \deduce{\Gamma_1 \vdash N_2}{\Pi_1}
  &
  \infer[L_1]
  {\Gamma_2 \vdash M'}
  {
    \deduce{\Gamma_2 \vdash N_1}{\Pi_2}
    &
    \deduce{\Gamma_4 \vdash M'}{\Pi_3'}
  }
}
$$
By inspection of the rules in Figure~\ref{fig:msg}, it can be shown
that this transformation is valid for any pair of left rules $(L_1,L_2).$
Note that this transformation may introduce at most two offending
left rules, i.e., if $\Pi_1$ and/or $\Pi_2$ end with left rules. 
But notice that the left premise derivations of both $L_1$ and $L_2$
in this case have smaller height than the left premise derivation 
of $L_1$ in $\Xi$. By induction hypothesis, the right premise
derivation of $L_2$ can be transformed into a normal derivation, say
$\Pi_4$, resulting in 
$$
\infer[L_2]
{\Gamma_1 \vdash M'}
{
  \deduce{\Gamma_1 \vdash N_2}{\Pi_1}
  &
  \deduce{\Gamma_2 \vdash M'}{\Pi_4}
}
$$
By another application of the induction hypothesis, this derivation
can be transformed into a normal derivation.  

The general case where $\Pi$ has more than one offending rules can be
dealt with by transforming the topmost occurrences of the left rule,
one by one, following the above transformation. 
\qed

In a normal derivation, the left branch of a branching left rule is
derivable using only right rules and $id$. This means that we can
represent a normal derivation as a sequence (reading the derivation bottom-up) 
of sequents, each of which is obtained from the previous
one by adding terms composed of subterms of the previous sequent, with
the proviso that certain subterms can be constructed using
right-rules.  Let us denote with $\Gamma \Vdash_\Rscr M$ the fact that
the sequent $\Gamma \vdash M$ is derivable using only the right rules
and $id$.  This suggests a more compact deduction system for intruder
deduction, called system $\Lscr$, given in Figure~\ref{fig:linear}.

\begin{figure}[t]
\center{
  \begin{tabular}[c]{ll}
   $\infer[r]{\Gamma \vdash M}
             {\Gamma \Vdash_\Rscr M}$
  &
   \qquad $\infer[le, \hbox{ where $\Gamma, \enc M K \Vdash_\Rscr K$}]
              {\Gamma, \enc M K \vdash N}
              {\Gamma, \enc M K, M, K \vdash N}$
\\[1em]
   $\infer[lp]{\Gamma, \spr M N \vdash T}
              {\Gamma, \spr M N, M, N \vdash T}$
   &
   \qquad $\infer[\signsym, K \equiv L]
              {\Gamma, \sign M K, \pub L \vdash N}
              {\Gamma, \sign M K, \pub L, M \vdash N}$
\\[1em]
  \multicolumn{2}{l}{
   $\infer[\blindsym_1, \hbox{ where $\Gamma, \blind M K \Vdash_\Rscr K$ }]
             {\Gamma, \blind M K \vdash N}
             {\Gamma, \blind M K, M, K \vdash N}$
  }
\\[1em]
  \multicolumn{2}{l}{
   $\infer[\blindsym_2,]
             {\Gamma, \sign {\blind M R} K \vdash N}
             {\Gamma, \sign {\blind M R} K, \sign M K, R \vdash N}$
  }
\\[1em] 
  \multicolumn{2}{l}{\qquad where $\Gamma, \sign {\blind M R} K \Vdash_\Rscr R.$} 
\\[1em]
  \multicolumn{2}{l}{
  $\infer[ls, 
           \hbox{where $A$ is an $E$-factor of $\Gamma \cup \{M\}$ 
                 and $\Gamma \Vdash_\Rscr A.$}]
            {\Gamma \vdash M}
            {\Gamma, A \vdash M} $
  }
  \end{tabular}
}
\caption{System $\Lscr$: a linear proof system for intruder deduction.}
\label{fig:linear}
\end{figure}

\begin{prop}
\label{prop:normal}
A sequent $\Gamma \vdash M$ is derivable in $\Sscr$ if and only if it
is derivable in $\Lscr.$
\end{prop}
\proof
This follows immediately from cut elimination for $\Sscr$ and the
normal form for $\Sscr$ (Proposition~\ref{prop:norm}). \qed

We now show that the decidability of the deduction problem $\Gamma \Vdash_\Sscr M$
can be reduced to decidability of elementary deduction problems. 
We consider a representation of terms as directed acyclic graphs (DAG), 
with maximum sharing of subterms. 
Such a representation is quite standard and can be found in, e.g., \cite{Abadi06TCS}, so
we will not go into the details here. 

In the following, we denote with $st(\Gamma)$ the set of subterms of the terms in $\Gamma.$
In the DAG representation of $\Gamma$, the number of distinct nodes in the DAG representing 
distinct subterms of $\Gamma$ co-incides with the cardinality of $st(\Gamma).$ 
We write $pst(\Gamma)$ for the set of proper subterms of $\Gamma$,
and write
$ St(\Gamma) $ 
for the {\em saturated set} of $\Gamma$, where
$$ St(\Gamma) = \Gamma \cup pst(\Gamma) \cup sst(\Gamma) \qquad
 sst(\Gamma) = \{\sign M N \mid M, N \in pst(\Gamma) \}$$
The set $sst(\Gamma)$ is needed so that the saturated set
is closed under the unblinding operation, i.e., the bottom-up
application of the $\blindsym_2$-rule.
The cardinality of $St(\Gamma)$ is at most quadratic in the size of
$st(\Gamma)$. If $\Gamma$ is represented as a DAG, one can 
compute the DAG representation of $St(\Gamma)$ in polynomial time, with only
a quadratic increase of the size of the graph. 
Given a DAG representation of $St(\Gamma \cup \{M\})$, 
we can represent a sequent $\Gamma \vdash M$ by associating 
each node in the DAG with a tag which indicates 
whether or not the term represented by the subgraph rooted at that node 
appears in $\Gamma$ or $M$. 
Therefore, in the following complexity results for the deducibility problem 
$\Gamma \Vdash_S M$ (for some proof system $S$), we assume that 
the input consists of the DAG representation of the saturated set 
$St(\Gamma \cup \{M\})$, together with approriate tags in the nodes.
Since each tag takes only a fixed amount of space (e.g., a two-bit data
structure should suffice), we shall state the complexity result
w.r.t. the cardinality of $St(\Gamma \cup \{M\}).$
We denote with $\#(\Sigma)$ the cardinality of the set $\Sigma.$

\begin{defi}
Let $\Gamma \Vdash_{\Dscr} M$ be a deduction
problem, where $\Dscr$ is some proof system, and let $n$ be the size of $St(\Gamma \cup \{M\}).$
Let $E$ be the equational theory associated with $\Dscr$. Suppose that
the elementary deduction problem in $E$ has complexity $O(f(m)),$ where $m$ is
the size of the input. 
Then the problem $\Gamma \Vdash_{\Dscr} M$ is said to be {\em polynomially reducible} 
to the elementary deduction problem $\Vdash_E$
if it has complexity $O(n^k \times f(n))$ for some constant $k.$
\end{defi}

A key lemma in proving the decidability result is 
the following invariant property of linear proofs.
\begin{lem}
\label{lm:St-inv}
Let $\Pi$ be an $\Lscr$-derivation of $\Gamma \vdash M.$
Then for every sequent $\Gamma' \vdash M'$ occurring in $\Pi$,
we have
$\Gamma' \cup \{M'\} \subseteq St(\Gamma \cup \{M\}).$
\end{lem}
\proof
By induction on $|\Pi|.$
It is enough to show that for each rule $\rho$ in $\Lscr$ other than $r$
$$
\infer[\rho]
{\Gamma \vdash M}
{\Gamma' \vdash M'}
$$
we have that $St(\Gamma \cup \{M\}) = St(\Gamma' \cup \{M'\})$. 

The non-trivial case is the rule $\blindsym_2$:
$$
\infer[\blindsym_2]
{\Gamma_1, \sign {\blind N R} K \vdash M}
{\Gamma_1, \sign {\blind N R} K, \sign N K, R \vdash M}
$$
where $\Gamma = \Gamma_1 \cup \{ \sign {\blind N R} K \}.$
The premise of the rule has a term $\sign N K$ which may not occur in the
conclusion. However, the proper subterms of $\sign N K$ are
included in the proper subterms of $\sign {\blind N R} K$, hence
both the premise and the conclusion have the same set of proper
subterms. 
Notice that $\sign N K \in sst(\Gamma)$, since
both $N$ and $K$ are in $pst(\Gamma).$ 
Therefore in this case we also have that 
$
St(\Gamma \cup \{ M \}) = St(\Gamma' \cup \{M'\}). 
$
\qed

The existence of linear size proofs then follows from the above
lemma.
\begin{lem}
\label{lm:linear}
If there is an $\Lscr$-derivation of $\Gamma \vdash M$ then there is an $\Lscr$-derivation
of the same sequent whose length is at most $\#(St(\Gamma \cup \{M\})).$
\end{lem}
\proof
  We first note that any derivation of $\Gamma \vdash M$ can be turned
  into one in which every sequent in the derivation occurs exactly
  once on a branch. Our rules preserve their principal formula when
  read upwards from conclusion to premise, hence the left hand sides
  of the sequents as we go up a branch accumulate more and more
  formulae. That is, they form an increasing chain. At worst, each
  such rule adds only one formula from $St(\Gamma \cup \{M\})$.  Thus,
  by Lemma~\ref{lm:St-inv}, the number of different sequents on a
  branch is bounded by the cardinality of $St(\Gamma \cup \{M\})$.
\qed

Another useful observation is that the left-rules of $\Lscr$ are {\em invertible};
at any point in a bottom-up proof search, we do not lose derivability by applying any 
left rule. Polynomial reducibility of $\Vdash_\Lscr$ to $\Vdash_E$ can
then be proved by a deterministic proof search strategy which systematically tries
all applicable rules. 

We now show that the decision problem $\Gamma \Vdash M$ is polynomially
reducible to the elementary deduction problem. 
This proof will make use of the linear proof system $\Lscr$. 
Since the side conditions in some rules in $\Lscr$ depend on $\Vdash_\Rscr$,
we first need to prove this reducibility result for $\Vdash_\Rscr.$ 
This is straightforward since the right introduction rules do not modify 
messages in the left hand side of the sequent, hence, if $m$ is the number
of distinct subterms of $M$, checking
this deducibility relation amounts to checking at most $m$
instances of $\Vdash_E$ on subterms of $M$.

\begin{lem}
\label{lm:right-deducibility}
The decidability of the relation $\Vdash_\Rscr$ is polynomially
reducible to the decidability of elementary deduction $\Vdash_E$.
\end{lem}
\proof
  Recall that the relation $\Gamma \Vdash_\Rscr M$ holds if we can
  derive $\Gamma \vdash M$ using only right-rules and $id$.  Here is a
  simple proof search procedure for $\Gamma \vdash M$, using only
  right-rules:
\begin{enumerate}[(1)]
\item If $\Gamma \vdash M$ is elementarily deducible, then we are
  done.
\item Otherwise, apply a right-introduction rule (backwards) to
  $\Gamma \vdash M$ and repeat step 1 for each obtained premise, and
  so on. If no such rules are applicable, then $\Gamma \vdash M$ is
  not derivable.
\end{enumerate}\medskip
There are at most $n$ iterations where $n$ is the number of distinct
subterms of $M.$ Note that the check for elementary deducibility in
step 1 is done on problems of size less or equal to $\#(St(\Gamma\cup M))$.  \qed

Before we proceed with proving the main decidability result 
(Theorem~\ref{thm:deducibility} below), let us first define the notion 
of a {\em principal term} in a  left-rule in the proof system $\Lscr$
(we refer to Figure~\ref{fig:linear} in the following definition):
\begin{enumerate}[$\bullet$]
\item $\spr M N$ is the principal term of $lp$
\item $\enc M K$ is the principal term of $le$
\item $\sign M K$ is the principal term of $\signsym$
\item $\blind M K$ is the principal term of $\blindsym_1$
\item $\sign {\blind M R} K$ is the principal term of $\blindsym_2$
\item $A$ is the principal term of $ls$.
\end{enumerate}\medskip
Given a sequent $\Gamma \vdash M$ and a pair of principal-term and a left-rule $(N, \rho)$, 
we say that the pair $(N,\rho)$ is {\em applicable} to the sequent if 
\begin{enumerate}[$\bullet$]
\item $\rho$ is $ls$, $N$ is a factor of $\Gamma \cup \{M\}$, and
there is an instance of $\rho$ with $\Gamma, N \vdash M$ as its premise;

\item $\rho$ is not $ls$, $N \in \Gamma$,  and there is an instance of $\rho$
with $\Gamma \vdash M$ as its conclusion.
\end{enumerate}

Let us assume that the complexity of $\Vdash_E$ is $O(f(n)).$
Given a sequent $\Gamma \vdash M$ and 
a pair of principal-term and a left-rule $(N, \rho)$, 
we note the following two facts:
\begin{enumerate}[\hbox to8 pt{\hfill}]
\item\noindent{\hskip-12 pt\bf F1:}\ the complexity of checking whether $(N, \rho)$ is applicable to $\Gamma \vdash M$
is $O(n^l f(n))$ for some constant $l$;

\item\noindent{\hskip-12 pt\bf F2:}\ if $(N, \rho)$ is applicable to $\Gamma \vdash M$, then there is a unique
sequent $\Gamma' \vdash M$ such that
the sequent below 
is a valid instance of $\rho$:
$$
\infer[\rho]
{\Gamma \vdash M}
{\Gamma' \vdash M}
$$
\end{enumerate}
Note that for (F1) to hold, we need to assume a DAG representation
of sequents with maximal sharing of subterms.
The complexity of checking whether a rule is applicable or not then consists of
\begin{enumerate}[$\bullet$]
\item pointer comparisons;
\item pattern matching a subgraph with a rule;
\item checking equality modulo associativity and commutativity (for the rule $\signsym$);
\item and checking $\Vdash_\Rscr$.
\end{enumerate}
The first three can be done in polynomial time; and the last one is polynomially
reducible to $\Vdash_E$ (Lemma~\ref{lm:right-deducibility}). 

\begin{thm}
\label{thm:deducibility}
The decidability of the relation $\Vdash_\Lscr$ is polynomially reducible
to the decidability of elementary deduction $\Vdash_E.$
\end{thm}
\proof
Let $n$ be the size of $St(\Gamma \cup \{M\})$.
Notice that the left-rules in Figure~\ref{fig:linear} are invertible 
(they accumulate terms, reading the rules bottom-up),
so one does not lose derivability by applying any of the rules in proof search.
Thus by blindly applying the left-rules, we
eventually reach a point where the right-rule ($r$) is applicable,
hence the original sequent is derivable, or we reach a ``fix point''
where we encounter all previous sequents.
For the latter, we show that there is a polynomial
bound to the number of rule applications we need to try
before concluding that the original sequent is not provable. 

Let $M_1, \ldots, M_n$ be an enumeration of the set $St(\Gamma \cup \{M\}).$
Suppose $\Gamma \vdash M$ is provable in $\Lscr$.
Then there is a shortest derivation 
in $\Gamma$ where each sequent appears exactly once in the (linear) derivation.
This also means that there exists a sequence of principal-term-and-rule pairs
$$
(M_{i_1}, \rho_1), \ldots, (M_{i_q}, \rho_q)
$$
that are applicable, successively, to $\Gamma \vdash M$.
Note that $q \leq n$ by Lemma~\ref{lm:linear}.

A simple proof search strategy for $\Gamma \vdash M$ is therefore to 
repeatedly try all possible applicable pairs $(M', \rho')$ for each possible
$M' \in St(\Gamma \cup \{M\})$ and each left-rule $\rho'$. More precisely:
Let $j := 0$ and initialise $\Delta := \Gamma$ 
\begin{enumerate}[(1)]
\item $j := j + 1$.
\item If $\Delta \Vdash_\Rscr M$ then we are done.
\item Otherwise, for $k = 1$ to $n$ do
  \begin{enumerate}[$\bullet$]
  \item[] for every left-rule $\rho$ do
    \begin{enumerate}[$-$]
  \item[] if $(M_k, \rho)$ is applicable to $\Delta \vdash M$, then
    let $\Gamma_1 \vdash M$ be the unique premise of $\rho$ determined
    by $(M_k, \rho)$ via {\bf F2} and let $\Delta := \Gamma_1$.
  \end{enumerate}
    \end{enumerate}
\item If $j \leq n$ then go to step 1.
\end{enumerate}
If the original sequent is derivable, then at each iteration $j$, 
the algorithm (i.e., step 3) will find the correct pair $(M_{i_j}, \rho_j)$.
(Strictly speaking, the algorithm finds 
the $j$-th pair of a shortest derivation,
and not necessarily the one given above, since there can be
more than one derivation of a given length.)
Note that the algorithm does not construct the shortest derivation,
but at each $j$ iteration, it will guess correctly the $j$-th pair of such a derivation
if one exists.  
If no derivation is found after $n$ (outer) iterations,
then the original sequent is not derivable,
since the length of any shortest derivation is bound by $n$ by 
Lemma~\ref{lm:linear}.
By Lemma~\ref{lm:right-deducibility},
step 2 takes $O(n^a f(n))$ for some constant $a$. 
By ({\bf F1}) above,
each iteration in step 3 takes $O(n^b f(n))$ for some constant $b$.
Since there are at most $6n$ distinct principal-term-and-rule pairs,
this means step 3 takes $O(6n^{b+1}f(n)).$ Therefore the whole procedure takes
$O(n^{c+1} f(n))$ where $c$ is the greater of $a$ and $b+1$.
Hence the complexity of $\Vdash_\Lscr$ is polynomially reducible to $\Vdash_E.$
\qed

Note that in the case where the theory $E$ is empty, we obtain a
\textsc{ptime} decision procedure for intruder deduction with blind
signatures.

\section{Combining disjoint convergent theories}
\label{sec:comb}

We now consider a slightly more general intruder deduction problem than the previous
sections: we shall allow any AC convergent theory which is obtained from a union 
of pairwise disjoint convergent AC theories. That is, the AC theory $E$ in
this case can be a disjoint combination of AC convergent theories 
$E_1, \ldots, E_n$, where each theory $E_i$ may contain an associative-commutative
binary operator, which we denote with $\oplus_i.$
We show that the intruder deduction problem
under $E$ can be reduced to the elementary deduction problem of each $E_i.$
The notions of subterms, factors, alien terms, etc., carry over to this more
general setting, but we shall be mostly concerned with the
constituent theories $E_i$'s, so we shall be speaking of $E_i$-alien terms,
$E_i$-factors, etc. 

The sequent system $\Sscr$ needs to be modified slightly to accomodate
this combination of theories. 
Throughout this section, we shall consider a sequent system $\Dscr$, whose rules
are those of $\Sscr$, but with $id$ replaced by 
the rule $id_{E_i}$ below left and
with the rule $acut$ below right:
$$
\infer[id_{E_i}]
{\Gamma \vdash M}
{
\begin{array}{c}
M \approx_{E} C[M_1,\ldots,M_k] \\
\hbox{$C[\ ]$ an $E_i$-context, and $M_1,\ldots,M_k \in \Gamma$}
\end{array}
}
\qquad \qquad
\infer[acut]
{\Gamma \vdash M}
{\Gamma \vdash N & \Gamma, N \vdash M}
$$
where $N$ is an $E_i$-factor of $\Gamma \cup \{M\}$. Notice that the sequent
system $\Sscr$ is then just a special case of $\Dscr$ where $E$ contains only 
a single AC operator. Note that in the proviso of the $id_{E_i}$ rule,
we require that $M \approx_{E} C[M_1,\ldots,M_k]$. However, as a consequence
Proposition~\ref{prop:var-abs} and Proposition~\ref{prop:var-abs2},
we have
$$
\begin{array}{ll}
M \approx_{E}  C[M_1, \ldots, M_k] &
\qquad
\mbox{iff}
\qquad C[F_E(M_1), \ldots, F_E(M_k)] \approx_E F_E(M) \\
& 
\qquad \mbox{iff} \qquad
C[F_E(M_1), \ldots, F_E(M_k)] \approx_{E_i} F_E(M).
\end{array}
$$
That is, in applying the $id_{E_i}$ rule, one can abstract all the $E_i$-alien
subterms from the sequent and check for equality in the theory $E_i$, rather than $E.$

A straightforward adaptation of the proof of Proposition~\ref{prop:S-equal-N}
gives an analog of it for $\Dscr$.

\begin{prop}
\label{prop:D-equal-N}
The judgment $\Gamma \vdash M$ is derivable in the natural deduction system $\Nscr$, under theory $E$, 
if and only if $\norm \Gamma \vdash \norm M$ is derivable in the sequent system $\Dscr$.
\end{prop}

Cut elimination also holds for $\Dscr$. Its proof is basically the same as the proof
for $\Sscr$, since the ``logical structures'' (i.e., those concerning constructors)
are the same. 
The crucial part of the proof in this case relies on the variable abstraction
technique (Proposition~\ref{prop:var-abs} and Proposition~\ref{prop:var-abs2}),
which applies to disjoint combination of theories. 
We can then prove the analog of the decomposition lemmas (Lemma~\ref{lm:decomp1} 
and Lemma~\ref{lm:decomp2}),
given below. 

\begin{lem}
\label{lm:comb-decomp1}
Let $X$ and $Y$ be terms in normal form and let $f$ be a binary
constructor.  If $\Gamma, f(X,Y) \vdash M$ is cut-free derivable, then
so is $\Gamma, X, Y \vdash M$.
\end{lem}
\proof
This is proved analogously to Lemma~\ref{lm:decomp1}. 
\qed

\begin{lem}
\label{lm:comb-decomp2}
Let $X_1,\ldots, X_k$ be normal terms and let 
$\Pi$ be a cut-free derivation of 
$$\Gamma, \norm{f(X_1,\ldots,X_k)} \vdash M,$$ 
where $f \in \Sigma_{E_i}.$
Then there exists a cut-free derivation $\Pi'$ of $\Gamma, X_1,\ldots,X_k \vdash M.$ 
\end{lem}
\proof
By induction on $|\Pi|.$
As in the proof of Lemma~\ref{lm:decomp2}, we do case analyses on the last rule of
$\Pi$. The cases involving constructors are the same as in the proof of Lemma~\ref{lm:decomp2}.
The non-trivial cases are when $\Pi$ ends with either $id$ or $acut$.

\begin{enumerate}[$\bullet$]
\item Suppose $\Pi$ ends with $id_{E_j}$: That is, we have
$$
C[\norm{f(X_1,\ldots,X_k)}\!\!\!^{n},M_1,\ldots,M_l] \approx M
$$
for some $E_j$-context $C[\ldots].$
If $i = j$ then $f \in \Sigma_{E_j}$ and the sequent
$\Gamma, X_1,\ldots,X_k \seqsym M$ is provable by an application of
$id_{E_j}$ using the $E_j$-context $C[f(\ldots)^n,\ldots].$

Otherwise, we have that $i \not = j$. 
Let $R = \norm{f(X_1,\ldots,X_k)}.$
There are two subcases to consider:
\begin{enumerate}[$-$]
\item $R$ is an $E_j$-alien term. Suppose $v(R) = x.$ Then by Proposition~\ref{prop:var-abs}
we have
$$
F_{E_j}(C[R^n,M_1,\ldots,M_l]) =
C[x^n, F_{E_j}(M_1),\ldots,F_{E_j}(M_l)] 
\rightarrow^* F_{E_j}(M).
$$
If $x$ does not occur in $F_{E_j}(M)$ then, using the same line of arguments 
as in the proof of Lemma~\ref{lm:decomp1}, it can be shown that
$$
C[X_1,M_1,\ldots,M_l] \approx M,
$$
hence $\Pi'$ in this case is a simple application of $id_{E_j}.$

Otherwise, if $x$ does occur in $F_{E_j}(M)$, 
then it can be shown that there exists $R' \equiv R$ such that either $R' = M$
or $R'$ is an $E_j$-factor of $M.$ For the former case, $\Pi'$ is
simply an application of the $id_{E_i}$ rule, since $f(X_1,\ldots,X_k) \approx M.$
For the latter case, we can apply the $acut$ rule to abstract $R'$ from $M$:
$$
\infer[acut]
{\Gamma,X_1,\ldots,X_k \seqsym M}
{
 \infer[id_{E_i}]
 {\Gamma,X_1,\ldots,X_k \seqsym R'}{f(X_1,\ldots,X_k) \approx R'}
 &
 \infer[id_{E_j}]
 {\Gamma, R', X_1,\ldots,X_k \seqsym M}
 {C[R'^n, M_1,\ldots,M_l] \approx M}
}
$$

\item $R$ is not an $E_j$-alien term, i.e., $R$ is headed by some $g \in E_j.$
This means that $R$ is an $E_i$-alien term. Since $f(X_1,\ldots,X_k) \rightarrow^* R$,
again using variable abstraction, it can be shown that there exists $R' \equiv R$ such
that either $R' \equiv X_p$ or $R'$ is an $E_i$-factor of $X_p.$
In either case, it is easy to construct a derivation of $\Gamma,X_1,\ldots,X_k \seqsym M.$
\end{enumerate}

\item Suppose $\Pi$ ends with $acut$
$$
\infer[acut]
{\Gamma,\norm{f(X_1,\ldots,X_k)} \seqsym M}
{
 \deduce{\Gamma, \norm{f(X_1,\ldots,X_k)} \seqsym A}{\Pi_1}
 &
 \deduce{\Gamma, \norm{f(X_1,\ldots,X_k)}, A \seqsym M}{\Pi_2}
}
$$
where $A$ is an $E_j$-factor of $\norm{f(X_1,\ldots,X_k)}.$
Note that $A$ in this case must be headed by a function symbol
not in $\Sigma_{E_j}.$

If $i = j$ then we have
$$
\norm{f(X_1,\ldots,X_k)} = C[g(\dots A \dots)]
$$
for some context $C[\dots]$ and some $g \in \Sigma_{E_i}.$
Again, using variable abstraction, it can be shown that there exists
$A' \equiv A$ and some $X_p$ such that either $A' = X_p$ 
or $A'$ is an $E_i$-factor of $X_p.$ For the former case, the derivation
$\Pi'$ is obtained by applying the induction hypothesis to $\Pi_2.$
For the latter case, the derivation $\Pi'$ is constructed as follows
$$
\infer[acut]
{\Gamma,X_1,\ldots,X_k\seqsym M}
{
  \deduce{\Gamma,X_1,\ldots,X_k \seqsym A'}{\Pi_1'}
  &
  \deduce{\Gamma,X_1,\ldots,X_k,A' \seqsym M}{\Pi_2'}
}
$$
where $\Pi_1'$ and $\Pi_2'$ are obtained from the induction hypothesis,
followed by applications of Lemma~\ref{lm:equiv}.

If $i \not = j$, then $g \not \in \Sigma_{E_i}$ and therefore
$g(\dots A \dots)$ is an $E_i$-alien term. 
In this case, there must exist $B \equiv g(\dots A \dots)$
such that $B$ is a subterm of some $X_p$. In other words,
$A$ is an $E_j$-factor of $X_p.$ So $\Pi'$ in this case
is constructed as in the derivation figure above. 
\qed
\end{enumerate}\medskip
We state the theorem below
and omit the proof since it is a straightforward adaptation of 
the cut elimination proof for $\Sscr.$

\begin{thm}
\label{thm cut elim for D}
The cut rule is admissible for $\Dscr$. 
\end{thm}
\proof
Analogous to the proof of Theorem~\ref{thm cut elim}, making use of Lemmas~\ref{lm:comb-decomp1}
and~\ref{lm:comb-decomp2}. \qed

The decidability result for $\Sscr$ also holds for $\Dscr.$ 
Its proof is basically the same as the decidability result for $\Sscr.$  
That is, we first show that derivations in $\Dscr$ admits
the same normal form as in $\Sscr.$
It then remains to design a linear proof system for $\Dscr.$ 
This is the same as $\Lscr$, except that the side condition of $ls$
is modified slightly: 
$$
\infer[ls]
{\Gamma \vdash M}
{\Gamma, N \vdash M} 
$$
where $N$ is an $E_i$-factor of $\Gamma \cup \{M\}$ and 
and $\Gamma \entail_{\Rscr} N$.  
We denote with $\Lscr\Dscr$ the linear proof system obtained
from $\Lscr$ by changing the $ls$ rule to the above one.
Then the following proposition is straightforward.

\begin{prop}
Every sequent $\Gamma \vdash M$ is derivable in $\Dscr$ if and only if it
is derivable in $\Lscr\Dscr$.
\end{prop}

The notion of polynomial reducibility is slightly changed. 
Suppose each elementary deduction problem in $E_i$ is bounded by $O(f(m)).$
Let $m$ be the size of $St(\Gamma \cup \{M\}).$ 
Then the deduction problem $\Gamma \Vdash_{\Dscr} M$ is polynomially reducible
to $\Vdash_{E_1}, \ldots, \Vdash_{E_n}$ if it has complexity
$O(m^k f(m))$, for some constant $k$.  
Note that here we only talk about the maximal complexity of the elementary
deduction in the {\em constituent theories}, and not 
the elementary deduction in the combined theory $E$, which may be higher.

\begin{thm}
\label{thm:D-deducibility}
The decidability of the relation $\entail_{\Lscr\Dscr}$ is polynomially reducible
to the decidability of elementary deductions $\Vdash_{E_1},$ $\ldots,$ $\Vdash_{E_n}$.
\end{thm}

\section{Deducibility constraints for Dolev-Yao intruders}
\label{sec:constraint}

We now consider a constraint problem that arises from
analysis of security protocols for a bounded 
number of sessions. This typically assumes
an active intruder which can synthesize messages from
a set of known messages, intercepted during runs of protocols,
to affect the running of the protocols. 
Since there could be infinitely 
many such messages, these need to be represented symbolically
as variables. As have been shown in a number of previous
works~\cite{Millen01,Boreale01ICALP,Comon-Lundh10ToCL}, the problem of finding 
an attack on a protocol for a bounded number of sessions 
(typically, violation of secrecy or authentication properties) 
can be mapped into the problem of solving {\em deducibility
constraints}. The latter are essentially a list of sequents,
possibly with occurrences of variables, and finding attacks to
a protocol then correspond to finding substitutions to the variables
such that the instances of the sequents under those substitutions
are derivable in the inference system modeling the intruder's abilities. 
We shall not delve into the specifics of the mapping from
protocol analysis into deducibility constraints; the interested
reader can consult the existing literature on the subject,
e.g., \cite{Millen01,Comon-Lundh10ToCL}. In this section, we report
on our preliminary study on how sequent calculus can be
applied to solve the deducibility constraint problem in
a limited setting, where the intruder model does not assume
any equational theories. For future work, we intend to study
the more general deducibility constraint problems involving
AC convergent theories. 

We note that the main results in this section have been
formally verified in the Isabelle/HOL proof assistant. 
The proof scripts are available via the web (given in
the introduction).

We shall be concerned only with Dolev-Yao intruders in
this section, i.e., we restrict to the constructors
$\spr . .$ and $\enc . .$, and an empty equational theory.
For this class of intruders, the deducibility constraint
problem has been shown decidable in 
several existing works~\cite{Boreale01ICALP,Millen01,Rusinowitch01CFSW,Comon-Lundh10ToCL}. 
In particular, our constraint reduction rules bear some similarity
with the reduction rules in \cite{Millen01}.
We shall, however, prove a stronger result, which is that
every deducibility constraint system is satisfiable 
if and only if it can be transformed into a certain solved form,
in which its solvability is immediate. 
A procedure for this transformation has been given recently in \cite{Comon-Lundh10ToCL}
using a natural deduction formulation of the intruder
model. Our aim here is to illustrate how the
sequent calculus can be used to solve the deducibility constraint
problem. 

Note that since we restrict to Dolev-Yao intruders, 
the rule $acut$ becomes redundant, since there could be no $E$-factors
in messages composed using constructors alone. 
Therefore in this case, the sequent system $\Sscr$ 
can be simplified to the one given in Figure~\ref{fig:dolev-yao}.

\begin{figure}
$$
\infer[id]{\Gamma \vdash M}
              {M \in \Gamma }
\qquad
\infer[p_L]{\Gamma, \spr M N \vdash T}
              {\Gamma, \spr M N, M, N \vdash T}
\qquad
\infer[p_R]
{\Gamma \vdash \spr M N}
{\Gamma \vdash M & \Gamma \vdash N}
$$

$$
\infer[e_L]
{\Gamma, \enc M K \vdash N}
{\Gamma, \enc M K \vdash K
   & \Gamma, \enc M K, M, K \vdash N}
\qquad
\infer[e_R]
{\Gamma \vdash \enc M K}
{\Gamma \vdash M & \Gamma \vdash K}
$$

\caption{Sequent system for Dolev-Yao intruders}
\label{fig:dolev-yao}
\end{figure}

\begin{defi}
\label{def:ded-constraint}
A {\em deducibility constraint} is an expression of the form
$\Sigma \entail^? M$ (called a {\em proper deducibility constraint}) 
or $\Sigma \entail^?_R M$ (called a {\em right-deducibility constraint}), where $\Sigma$ is
a set of messages and $M$ is a message. $\Sigma$ here is
called the left side of the constraint and $M$ its right side.
We write $\Sigma \entail^?_{(R)} M$
to denote a constraint generally without referring to its specific
form. 
\end{defi}
Intuitively, the constraint $\Sigma \entail^? M$ denotes
the problem of finding a derivable instance of the sequent
$\Sigma \seqsym M$, 
while the constraint $\Sigma \entail^?_R M$ denotes the problem of finding
an instance of the sequent $\Sigma \seqsym M$ that is derivable
using only the identity and the right-rules. 
The separation of constraints into these two kinds is motivated
by the structure of normal derivations, which separates
proof search into general deducibility and right-deducibility. Indeed,
our decision procedure for solving constraints exploits the 
structure of normal derivations.

If $C$ is a list of constraints, then $V(C)$ denotes the
set of variables occuring in $C$.
A {\em substitution} is a mapping from variables to terms.
It is extended to a mapping from terms to terms in the usual way.
We denote with $dom(\theta)$ the {\em domain} of the substitution $\theta$,
and $ran(\theta)$ denotes its range. 
We denote with $\epsilon$ the substitution with empty domain, i.e.,
the identity map on variables. 
A substitution $\theta$ is a {\em ground substitution} if
$\theta(x)$ is a ground message for every $x \in dom(\theta).$ 
Application of a substitution $\theta$ to a message $M$ is written
in a postfix notation, i.e., $M\theta$. This notation generalises
to sets of terms, sequents, constraints, etc., in the obvious way, 
e.g., $\Gamma\theta$ denotes the set of
messages obtained from applying the substitution $\theta$ to
each member of the set. Composition of substitutions is written
$\theta \circ \rho$ and is defined as $M(\theta \circ \rho) = (M\theta)\rho.$

\begin{defi} \label{def:solution,satisfiable}
A ground substitution $\theta$ is a 
{\em solution to a list of deducibility constraints $C$} if 
\begin{enumerate}[$\bullet$]
\item for every $\Sigma \entail^? M \in C$,
we have $\Sigma\theta \entail M \theta$, and 
\item for every $\Sigma \entail^?_R M \in C$,
we have $\Sigma\theta \entail_R M\theta$.
\end{enumerate}\medskip
We say that $C$ is {\em satisfiable} if there is a solution for $C$.
\end{defi}

Given a list of constraints $C$ and an index $i$, we write
$C^i$ to denote the prefix of $C$ of length $(i-1).$
So, if $C$ is, for example, 
$$
(\Sigma_1 \Vdash^? M_1) ; (\Sigma_2 \Vdash^? M_2) ; (\Sigma_3 \Vdash^? M_3)
$$
then $C^1$ is the empty list; $C^2$ is the singleton list
$(\Sigma_1 \Vdash^? M_1).$
Obviously, if $\theta$ is a solution for $C$ then
it is also a solution for any of its prefixes.

In the following, given $\Sigma_1$ and $\Sigma_2$, we write
$\Sigma_1 \entail \Sigma_2$ if $\Sigma_1 \entail M$ for
every $M \in \Sigma_2.$

\begin{defi}
\label{def:ded-constraint-system-alt}
A {\em deducibility constraint system} $C$ is a list of deducibility constraints
$$
\Sigma_1 \entail^?_{(R)} M_1 ; \cdots ; \Sigma_n \entail^?_{(R)} M_n
$$
such that:
\begin{enumerate}[(1)]
\item \label{lhs-dycond}
For $i < j$ if $\Sigma_j^{dv}$ is obtained from $\Sigma_j$ by deleting
messages which contain a variable not in any message in $\Sigma_i$,
then for all solutions $\theta$ to $C^j$, 
$\Sigma_j^{dv} \theta \entail \Sigma_i \theta$.

\item \label{dcs-fnrf} 
For every variable $x \in V(C)$, there exists 
$\Sigma_i \entail^?_{(R)} M_i$ such that $x \in V(M_i)$,
$x \not \in V(\Sigma_i)$, and for every $j < i$,
$x \not \in V(\Sigma_j \entail^?_{(R)} M_j).$
The index $i$ in this case is called the {\em order of $x$}
and will be denoted by $Ord(x)$.

\end{enumerate}
\end{defi}

\begin{rem}
\label{rem:monotonicity}
A commonly used definition of deducibility constraint systems
(in the natural-deduction-based approach) imposes a condition 
that the lefthand sides of the constraints (the $\Sigma_i$'s)
are ordered by set inclusion (see e.g., \cite{Rusinowitch01CFSW,Comon-Lundh10ToCL}).
This condition captures the fact that the knowledge of the intruder
increases with time as it accumulates more messages. 
Our definition of a deducibility constraint system is slightly
different in this respect. We capture this monotonicity condition
via the deduction relation itself. This is somewhat more complicated
than the natural deduction counterpart, but it is essentially imposed
by our choice of the reduction rules on constraints: a natural choice
of the reduction rules is one which mimics closely the inference rules
of the proof system, hence we allow decomposition of messages 
on both the lefthand sides and the righthand sides of
constraints, in contrast to the natural-deduction-based approach where
decomposition of messages happens only on the righthand sides. 
Note that in Condition~\ref{lhs-dycond} in Definition~\ref{def:ded-constraint-system-alt},
if the lefthand sides of the constraints are totally ordered by
set inclusion, then $\Sigma^{dv}_j \supseteq \Sigma_i$, hence trivially,
$\Sigma_j^{dv}\theta \entail \Sigma_i\theta$. Therefore, our definition
of deducibility constraint system subsumes that used in
the natural-deduction-based approaches.
\end{rem}

\begin{defi} \label{def:solved-form}  
A deducibility constraint system $C$ is in {\em solved form} if every element
in $C$ is of the form $\Sigma \entail^?_R x$ for some $\Sigma$ and variable $x.$
\end{defi}

For simplicity, we shall assume that in a deducibility constraint system $C$
$$
\Sigma_1 \entail^?_{(R)} M_1 ; \cdots ; \Sigma_n \entail^?_{(R)} M_n
$$
there is a name, say $a$, that is in every $\Sigma_i.$ 
As a consequence, if $C$ is in solved form, then it is trivially
solvable: simply instantiate every variable in $V(C)$ to $a$. 
This assumption is harmless as far as reasoning about protocols is concerned,
since in this setting, the intruder is usually assumed to have access to
infinitely many ``environment'' names. Some work in the literature,
e.g., \cite{Boreale01ICALP}, chooses to make this explicit by 
adding a special inference rule for deriving environment names. 

The goal of this section is to show that every deducibility constraint
system can be transformed into a deducibility constraint system in solved form,
preserving the set of solutions. 

\begin{defi}
\label{def:constraints-rewrite-rules}
The family of relations $\cred \theta$, where $\theta$ is a substitution, relate
lists of constraints and are defined below. 
If $\theta$ is the identity substitution we write $\credid$ instead of $\cred\theta.$
\begin{enumerate}[\hbox to8 pt{\hfill}]
\item\noindent{\hskip-12 pt\bf C1:}\ $C_1 ; \Sigma \entail^?_R M ; C_2 \cred\theta C_1\theta ; C_2\theta$, if 
$M$ is not a variable and there exists $N \in \Sigma$ such that $\theta = mgu(M, N).$

\item\noindent{\hskip-12 pt\bf C2:}\ $C_1 ; \Sigma \entail^?_R f(M,N) ; C_2 \credid C_1 ; \Sigma \entail^?_R M 
; \Sigma \entail^?_R N ; C_2$, where $f$ is either $\spr . .$ or $\enc . .$.

\item\noindent{\hskip-12 pt\bf C3:}\ $C_1 ; \Sigma \entail^? M ; C_2 \credid C_1 ;  \Sigma \entail^?_R M ; C_2$.

\item\noindent{\hskip-12 pt\bf C4:}\ $C_1 ; (\Sigma, \spr M N \entail^? U) ; C_2 
\credid C_1 ; (\Sigma, M, N \entail^? U) ; C_2$,
where $\spr M N \not\in \Sigma$.
\item\noindent{\hskip-12 pt\bf C5:}\ $C_1 ; (\Sigma, \enc M N
  \!\entail^?\!U) ; C_2 \credid C_1 ; (\Sigma, \enc M N \!\entail^?_R\!N) ;
  (\Sigma, M, N \!\entail^?\!U) ; C_2 $, where $\enc M N \not\in
  \Sigma$.
\end{enumerate}

\end{defi}

Notice that in {\bf C4} and {\bf C5}, when $M$ and $N$ are already in $\Sigma$,
then these steps are essentially a weakening step, as they remove a pair or
an encrypted message from the lefthand side of a constraint. 
Notice also that the reduction is defined on lists of constraints, not just
constraint systems. But as we shall see later, the reduction does preserve
the property of being a deducibility constraint system. This preservation
will be used in proving the completeness of the reduction rules for
deducibility constraint systems.

\begin{lem}[Soundness] 
\label{lm:constraint-red-sound}
Let $C$ be a list of constraints and suppose $C \cred\theta C'.$
If $C'$ is solvable then $C$ is also solvable.
Moreover, if $\sigma$ is a solution for $C'$ then
$\theta \circ \sigma$ is a solution for $C.$
\end{lem}
\proof
The reduction rules {\bf C1} to {\bf C3} are obviously sound
({\bf C1} relies on the properties of mgu).
For {\bf C4} and {\bf C5}, we need to apply 
the weakening lemma (Lemma~\ref{lm:weak}). 
\qed

An immediate consequence of Lemma~\ref{lm:constraint-red-sound}
is that, if $C$ rewrites to a solved form, then 
$C$ is satisfiable, and a solution for $C$ can be computed by
composing the substitutions associated with the reduction. 

\begin{lem} \label{lm:larger-reducible} 
If $C_1 ; \Sigma \entail^? M ; C_2$ is reducible,
and $\Sigma \subseteq \Sigma'$
then $C_1 ; \Sigma' \entail^? M ; C_2$ is reducible.
\end{lem}

\begin{lem} \label{lm:red-pres-dcs}
If $C$ is a deducibility constraint system and $C \cred \theta C'$ then 
$C'$ is also a deducibility constraint system.
\end{lem}
\proof
Condition~\ref{lhs-dycond} of Definition~\ref{def:ded-constraint-system-alt}
requires that, for constraints 
$ \Sigma_i \entail^?_{(R)} X_i $ and $ \Sigma_j \entail^?_{(R)} X_j $,
for all solutions $\sigma$ of $C^j$, 
$ \Sigma_j^{dv} \sigma \entail \Sigma_i \sigma $,
where $ \Sigma_j^{dv}$ is $ \Sigma_j$, modified by deleting messages containing
variables which are not in $\Sigma_i$.

We first note that this property is preserved by a substitution which 
arises in the reduction rule {\bf C1}.
Suppose $C \cred\theta C'$ by rule {\bf C1},
and let $\sigma$ be a solution for $C'$.
Then, by Lemma~\ref{lm:constraint-red-sound},
$\theta \circ \sigma$ is a solution for $C$, hence also a solution for $C^j$.
So we have 
$\Sigma_j^{dv} (\theta \circ \sigma) \entail \Sigma_i (\theta \circ \sigma)$,
and we require 
$(\Sigma_j\theta)^{dv} \sigma \entail (\Sigma_i\theta) \sigma$,
where $(\Sigma_j\theta)^{dv}$ is obtained by removing from $\Sigma_j\theta$
messages containing variables which are not in $\Sigma_i\theta$.
But if $M\theta$ is such a message, then $M$ must contain variables which are
not in $\Sigma_i$, and so $M$ has been removed in constructing $\Sigma_j^{dv}$
from $\Sigma_j$.
Therefore $\Sigma_j^{dv} \theta \subseteq (\Sigma_j\theta)^{dv}$
and so 
$\Sigma_j^{dv} (\theta \circ \sigma) \entail \Sigma_i (\theta \circ \sigma)$
implies
$(\Sigma_j\theta)^{dv} \circ \sigma \entail (\Sigma_i\theta) \sigma$.

Reduction rules {\bf C2} and {\bf C3} do not change the left-hand side of a
constraint, so the only issue they raise is that {\bf C2} produces two
constraints from one --- this gives an additional case of constraints
$ \Sigma_i \entail^?_{(R)} X_i $ and $ \Sigma_j \entail^?_{(R)} X_j $.
However here, $ \Sigma_i = \Sigma_j$ which satisfies this requirement.

Reduction rule {\bf C4}: Consider the requirement that
$ \Sigma_j^{dv} \sigma \entail \Sigma_i \sigma $.
If $\Sigma_i$ is changed to $\Sigma_i'$ by an application of rule {\bf C4},
then we have $ \Sigma_i \sigma \entail \Sigma_i' \sigma $
and so $ \Sigma_j^{dv} \sigma \entail \Sigma_i' \sigma $.
(It is also necessary to observe that $\Sigma_i'$ contains the same variables
as does $\Sigma_i$, and so $ \Sigma_j^{dv}$, defined relative to $\Sigma_i'$,
is the same as $ \Sigma_j^{dv}$, defined relative to $\Sigma_i$).
If $\Sigma_j$ is changed to $\Sigma_j'$ by an application of rule {\bf C4},
then we have $ \Sigma_j' \sigma \entail \Sigma_j \sigma $.
Further, note that when, say, $\Sigma_j = \Omega, \spr M N$,
and $\Sigma_j' = \Omega, M, N$,
if either $M$ or $N$ is deleted in forming $\Sigma_j'^{dv}$, then 
$\spr M N$ is deleted in forming $\Sigma_j^{dv}$.
Thus we get $ \Sigma_j'^{dv} \sigma \entail \Sigma_j^{dv} \sigma $
and so $ \Sigma_j'^{dv} \sigma \entail \Sigma_i \sigma $.

Reduction rule {\bf C5}: In part, the argument is similar to that for {\bf C4}.
If $\Sigma_i$ is subject to an application of rule {\bf C5},
say $\Sigma_i = \Omega, \enc M N$
then the first new constraint resulting is $\Omega, \enc M N \entail^?_R N$,
which has the same left-hand side.
The second new constraint resulting is $\Omega, M N \entail^?_R X_i$,
and we have that if $\sigma$ is a solution of $C'$ (and so 
$(\Omega, \enc M N)\sigma \entail N\sigma$)
then we get $(\Omega, \enc M N)\sigma \entail (\Omega, M N)\sigma$,
and so $\Sigma_i^{dv}\sigma \entail (\Omega, M N)\sigma$, as required.

If $\Sigma_j$ is subject to an application of rule {\bf C5},
then the argument is similar to that for rule {\bf C4}.

Finally if we consider the two constraints resulting from rule {\bf C5},
it is easy to check that the condition holds.  

Condition~\ref{dcs-fnrf} of Definition~\ref{def:ded-constraint-system-alt}
is that any variable appears on the right-hand side of a
constraint before it appears on the left-hand side of any constraint
(equivalently, any variable which in the left-hand side of any
constraint appears in an earlier constraint).

We first show that this property is preserved by any substitution.
Consider a constraint system 
$ \Sigma_1 \entail^?_{(R)} X_1 ; \cdots ; \Sigma_n \entail^?_{(R)} X_n $
and a substitution $\theta$.
Let $x$ be in $\Sigma_k\theta$.
Then for some $y$ in $\Sigma_k$, $x$ is in $y\theta$.
Now as $y$ must be in some earlier $X_j$ ($j < k$),
$x$ is in $X_j\theta$, as required.
Reduction {\bf C1} consists of a substitution, then deleting a constraint
$\Sigma \entail_R^? M$ for which $M \in \Sigma$. Clearly deleting such a constraint also preserves 
condition~\ref{dcs-fnrf} of Definition~\ref{def:ded-constraint-system-alt}.

It is straightforward to check that condition~\ref{dcs-fnrf} is preserved
by reductions {\bf C2} to {\bf C5}.
\qed

Given a term $M$, we denote by $|M|$ the size of the term $M$.
Given a set of terms $\Sigma$, define $|\Sigma| = \sum_{M \in \Sigma} |M|.$

\begin{defi}
\label{def:constraints-measure}
Let $\Sigma$ be a set of messages. 
We define a measure on deducibility constraints, 
denoted by $|\cdot|$ as follows:
$$
|\Sigma \entail^?_R M| = (0, |M|) \qquad
|\Sigma \entail^? M| = (1, |\Sigma|)
$$
Deducibility constraints are ordered by lexicographical ordering
on their measures. 

The measure of a deducibility constraint system $C$, denoted by $|C|$, is
$$
|C| = (\# V(C), S)
$$
where $S$ is the multiset of measures of the deducibility constraints
in $C.$ 
There is a well-founded ordering on constraints systems, i.e.,
one which is obtained by lexicographical ordering on $|C|$, 
where the first component is ordered according to $\leq$ on natural
numbers, and the second component is ordered according to
multiset ordering (parameterised on the ordering on deducibility constraints). 
\end{defi}

\begin{lem}[Termination of constraints reduction]
\label{lm:constraint-red-terminates}
For every constraint system $C$, there is no infinite reduction
sequence starting from $C.$
\end{lem}
\proof
It is enough to show that each instance of the rewrite rules {\bf C1} to {\bf C5}
reduces the measure on constraint systems. That is,
we show that whenever $C \cred\theta C'$ then $|C'| < |C|.$
For {\bf C1}, by the properties of mgu, the number of variables in $C'$ is smaller than 
or equal to the number of variables in $C$, but the number of deducibility
constraints in $C'$ is smaller than $C$, so $|C'| < |C|$. 
All other cases are straightforward from Definition~\ref{def:constraints-rewrite-rules} 
and Definition~\ref{def:constraints-measure}.
\qed

In the following, a rewrite sequence such as 
$$
C_1 \cred{\theta_1} C_2 \cred{\theta_2} \cdots \cred{\theta_{n-1}} C_n
$$
shall be abbreviated as $C_1 \credtr \theta C_n$ where
$\theta = \theta_1 \circ \cdots \circ \theta_{n-1}.$
Given two substitutions $\theta$ and $\sigma$, and a set of variables $V$,
we write 
$$
\theta =_V \sigma
$$
when $\theta$ and $\sigma$ coincide on $V.$

\begin{lem}[Completeness] 
\label{lm:constraint-red-complete}
Let $C$ be a constraint system and let $\theta$ be a solution for $C.$
Then there exists a rewrite sequence $C \credtr \sigma C'$
such that $C'$ is in solved form, 
$\theta =_{V(C)} \sigma \circ \gamma,$ 
and $\gamma$ is a solution for $C'.$
\end{lem}
\proof
We prove this by induction on $|C|.$
If $C$ is in solved form (this includes the case where $C$ is empty), 
then let $C' = C$ and let $\gamma = \theta$ and $\sigma$ be
the identity substitution. Otherwise, since $\theta$ is a solution for $C$, 
for every $\Sigma \entail^?_{(R)} M \in C$, we have
$\Sigma\theta \entail_{(R)} M\theta.$ Without loss of generality, we assume
that all derivations are in normal form. 
We construct a rewriting sequence
on $C$ by examining the last rule of a selected constraint in $C.$

By definition, elements of $C$ can be listed as
$$
\Sigma_1 \entail^?_{(R)} M_1 ; \cdots ; \Sigma_n \entail^?_{(R)} M_n
$$
Let $i$ be the maximal index such that $C^i$ is in solved form.
We shall select the constraint $\Sigma_i \entail^?_{(R)} M_i$ 
as a candidate for reduction. 

We now proceed to showing that it is always possible to apply  
a rewrite rule to the selected constraint such that $C \cred\rho D$, for 
some constraint system $D$, and such that $\theta = \rho \circ \beta,$
and $\beta$ is a solution of $D$. 
There are several possible rewritings on the selected constraint, depending
on the last rule of the normal derivation of the selected constraint:
\begin{enumerate}[(1)]

\item Suppose the selected constraint is a right-deducibility constraint,
and suppose that there is a normal derivation of 
$\Sigma_i\theta \seqsym M_i\theta$ ending with an $id$.
That is, $M_i\theta = N\theta$ for some $N \in \Sigma_i.$
Let $\rho = mgu(M_i, N)$. 
Then rewrite $C$ using {\bf C1}:
$$
C = C^i ; (\Sigma_i \entail^?_R M_i) ; C_1  \cred\rho C^i\rho ; C_1\rho = D
$$
Obviously, $\theta = \rho \circ \beta$ for
some $\beta$, and $\beta$ is a solution to $D.$

\item Suppose the selected constraint is $\Sigma_i \entail^?_R f(M,N)$,
where $f$ is either $\spr . .$ or $\enc . .$, 
and the normal derivation of $\Sigma_i\theta \seqsym f(M\theta,N\theta)$
ends with a right-introduction rule. The latter means that
$\Sigma_i\theta \entail_R M\theta$ and $\Sigma_i\theta \entail_R N\theta.$
Then rewrite $C$ using {\bf C2}:
$$
C = C^i ;  (\Sigma_i \entail^?_R f(M,N)) ; C_1 
\credid C^i ; (\Sigma \entail^?_R M) ; (\Sigma \entail^?_R N) ; C_1 = D.
$$
Obviously, $\theta$ is also a solution to $D$, so in this case,
$\rho = \epsilon$ and $\beta=\theta.$

\item Suppose the selected constraint is $\Sigma_i \entail^? M_i$ but
the normal derivation of $\Sigma_i\theta \seqsym M_i\theta$ ends with a right-rule.
The latter means that $\Sigma_i \theta \entail_R M_i\theta.$
Then rewrite $C$ using {\bf C3}:
$$
C = C^i ; (\Sigma_i \entail^? M_i) ; C_1 
\credid C^i ; (\Sigma_i \entail^?_R M_i) ; C_1 = D.
$$
Obviously, $\theta$ is also a solution to $D$, so $\rho = \epsilon$
and $\beta = \theta.$

\item Suppose the selected constraint is a proper deducibility constraint
and suppose there exists $M \in \Sigma_i$, i.e., $\Sigma_i = \Sigma_i'\cup \{M\}$, 
such that $M$ is not a variable, and there is a normal derivation of $\Sigma_i'\theta, M\theta \seqsym M_i\theta$ ending 
with a left rule applied to $M\theta$. 
Since $M$ is not a variable, it must be either a pair $\spr {N_1} {N_2}$ or
an encrypted term $\enc {N_1} {N_2}.$
\begin{enumerate}[$\bullet$]
\item If $M = \spr {N_1} {N_2}$, then, by normal derivability of
$\Sigma_i' \theta, M\theta \seqsym M_i\theta$, we have that
$$
\Sigma_i'\theta, M\theta, N_1\theta, N_2\theta \entail M_i\theta.
$$
Note that by Lemma~\ref{lm:decomp1}, we also have
$$
\Sigma_i'\theta, N_1\theta, N_2\theta \entail M_i\theta.
$$

In this case, apply the rewrite rule {\bf C4}:
$$
C = C^i ; (\Sigma_i', M \entail^? M_i) ; C_1 \credid 
C^i ; (\Sigma_i', N_1, N_2 \entail^? M_i) ; C_1 = D.
$$
Then $\theta$ is obviously a solution for $D$.
As in the previous case, let $\rho = \epsilon$ and $\beta = \theta.$

\item If $M = \enc {N_1}{N_2}$ then we have
$$
\Sigma_i'\theta, M \theta \entail_R N_2\theta
\quad \hbox{ and } \quad
\Sigma_i'\theta, M\theta, N_1\theta, N_2\theta \entail M_i\theta.
$$
By Lemma~\ref{lm:decomp1}, we also have
$$
\Sigma_i'\theta, N_1\theta, N_2\theta \entail M_i\theta.
$$
In this case, apply the rewrite rule {\bf C5}:
$$
C = C^i ; (\Sigma_i', M \entail^? M_i) ; C_1 
\credid C^i ; (\Sigma_i', M \entail^?_R N_2) ; 
(\Sigma_i', N_1, N_2 \entail^? M_i) = D.
$$
It is clear that $\theta$ is also a solution to $D$,
so let $\rho = \epsilon$ and $\beta = \theta.$
\end{enumerate}\medskip
Note that in both cases, Lemma~\ref{lm:decomp1} does not need
to be applied if $M\theta \in \Sigma_i'\theta$, since in this case
we have
$$
(\Sigma_i' \cup \{N_1,N_2\})\theta = 
(\Sigma_i' \cup \{M, N_1, N_2\})\theta.
$$

\item Suppose the selected constraint is 
$$\Sigma_i', x_{1},\ldots,x_{n} \entail^? M_i$$ 
where $\Sigma_i'$ contains only non-variable terms. 
Note that since $C^i$ is in solved form, and since $C$
is a deducibility constraint system, it must be the case that each $x_k$ appears
in the righthand side of a constraint in $C^i$.%
\footnote{More precisely, since $C$
is a deducibility constraint system, it must be the case that each $x_k$ appears
in the righthand side of a constraint in $C^i$,
and since $C^i$ is in solved form, each $x_k$ is
the righthand side of a constraint in $C^i$.}
Obviously, any two distinct variables $x_k$ and $x_l$ cannot 
be the same righthand side, therefore, without loss of 
generality, we assume that $Ord(x_k) < Ord(x_l)$
whenever $k < l.$ Notice that by well-formedness of $C$,
$Ord(x_l) < i$ for every $l \in \{1,\ldots,n\}.$

Suppose that there is a normal derivation $\Pi$ of the sequent
\begin{equation}
\label{eq:comp0}
\Sigma_i'\theta, x_{1}\theta, \ldots, x_{n}\theta \seqsym M_i\theta
\end{equation}
which ends with a left rule applied to one of $x_{k}\theta.$
We first show that the following sequent is derivable
\begin{equation}
\label{eq:comp1}
\Sigma_i'\theta \seqsym M_i\theta.
\end{equation}
To derive the above sequent, we first note the following facts:
\begin{enumerate}
\item Since $C^i$ is in solved form, we have for each $k \in \{1,\dots,n\}$, 
$(\Sigma_{o(k)} \entail^?_{R} x_k) \in C^i$, where $o(k)$ is the order of $x_k$,
hence 
\begin{equation}
\label{eq:comp2a}
\Sigma_{o(k)}\theta \entail_{R} x_k\theta.
\end{equation}

\item Let $\Sigma^k_i = \Sigma_i' \cup \{x_1,\ldots,x_{k-1}\}$ for $k \leq n.$
Since $C$ is a deducibility constraint system, by Definition~\ref{def:ded-constraint-system-alt}(\ref{lhs-dycond}), 
there exists $\Omega_{k} \subseteq \Sigma^k_i$ such that
$V(\Omega_{k}) \subseteq V(\Sigma_{o(k)})$ and
$$
\Omega_{k}\theta \entail \Sigma_{o(k)}\theta
$$
by definition, hence by weakening (Lemma~\ref{lm:weak}),
$
\Sigma_i^k\theta \entail \Sigma_{o(k)}\theta. 
$
Then by several applications of cut (using Sequent~(\ref{eq:comp2a}) above),
we get
\begin{equation}
\label{eq:comp2b}
\Sigma^k_i\theta \entail x_k\theta
\end{equation}
for any $k \leq n.$
\end{enumerate}\medskip
Applying cuts successively using instances of 
Sequent~(\ref{eq:comp2b}) and Sequent~(\ref{eq:comp0}),
we obtain Sequent~(\ref{eq:comp1}) as required.

Then consider a normal derivation of Sequent~(\ref{eq:comp1}).
The arguments of the previous cases show that the constraint 
$\Sigma_i' \entail^? M_i$ would admit a reduction.  
It follows trivially (similarly to Lemma~\ref{lm:larger-reducible})
that the enlarged sequent 
$\Sigma_i', x_{1},\ldots,x_{n} \entail^? M_i$ would admit a reduction.
\end{enumerate}\medskip
Since rewriting reduces the size of the constraint system,
by induction hypothesis
$D \credtr{\rho'} C'$ such that $C'$ is in solved form, 
$\beta =_{V(D)} \rho' \circ \gamma'$ and $\gamma'$ is a solution for $C'.$
Now let $\sigma = \rho \circ \rho'$ and let $\gamma = \gamma'$.
Then we indeed have $C \credtr \sigma C'$,  
$\theta =_{V(C)} \sigma \circ \gamma$ and $\gamma$ is a solution for $C'.$
\qed

\begin{thm}[Decidability of deducibility constraints]
\label{thm:dec-constraints}
Given a deducibility constraint system $C$, it is decidable whether or not
the constraint is satisfiable. 
\end{thm}
\proof
This is a consequence of Lemma~\ref{lm:constraint-red-terminates}, 
Lemma~\ref{lm:constraint-red-sound}, Lemma~\ref{lm:constraint-red-complete}
and the fact that the rewrite system $\leadsto$ is finitely branching.
\qed

To conclude this section, we shall comment briefly on the main
differences between our approach and that of Comon-Lundh, et. al., \cite{Comon-Lundh10ToCL}.
Apart from the difference in the way we impose the monotonicity
condition (see Remark~\ref{rem:monotonicity}), the main difference
is of course in the reduction rules.\footnote{They also consider
a slightly richer intruder model, containing asymmetric encryption and signing.
But it is easy to extend our work to accomodate these additional operators.} 
In their work, no explicit decomposition is applied to the left-hand side 
of a constraint. Instead, they allow unification of arbitrary subterms in a constraint. 
Our reduction rules, on the other hand, have a direct correspondence with
the inference rules of the proof system itself. This could perhaps be beneficial
when dealing with theories for which the subformula property does not hold,
e.g., when it involves blind signatures, where exhaustive unification tests 
on subterms may not be sufficient to get completeness.

\section{Conclusion and related work}
\label{sec:rel}

We have shown that decidability of the intruder deduction problem, under a range
of equational theories, can be reduced to the simpler problem
of elementary deduction, which amounts to solving equations in the underlying
equational theories. 
In particular, this reduction is obtained in a purely proof theoretical way,
using standard techniques such as cut elimination and
permutation of inference rules. We show that sequent-based techniques
can also be used to solve the deducibility constraint problems, for Dolev-Yao
intruders. 

There are several 
existing works in the literature that deal with intruder 
deduction. 
Our work is more closely related to, e.g.,
\cite{Comon-Lundh03LICS,Delaune06IPL,Lafourcade07IC}, in that we do
not have explicit destructors (projection, decryption, unblinding),
than, say, \cite{Abadi06TCS,Cortier07LPAR}. In the latter work, these
destructors are considered part of the equational theory, so in this
sense our work slightly extends theirs to allow combinations of
explicit and implicit destructors.  A drawback for the approach with
explicit destructors is that one needs to consider these destructors
together with other algebraic properties in proving decidability,
although recent work in combining decidable theories
\cite{Arnaud07frocos} allows one to deal with them modularly.
Combination of intruder theories has been considered in
\cite{Chevalier05ICALP,Arnaud07frocos,DelauneIC08}, as part of
their solution to a more difficult problem of deducibility constraints
which assumes active intruders.  In particular,
Delaune, et. al., \cite{DelauneIC08} obtain results similar to what we
have here concerning combination of AC theories.
One difference between these works and ours is in how this
combination is derived.  Their approach is more algorithmic whereas
our result is obtained through analysis of proof systems.

It remains to be seen whether sequent calculus, 
and its associated proof techniques, can prove useful for richer theories.
For certain deduction problems,
i.e., those in which the constructors interact with the equational theory,
there do not seem to be general results
like the ones we obtain for theories with no interaction with the constructors.
One natural problem where this interaction occurs is the theory
with homomorphic encryption,
e.g., like the one considered in \cite{Lafourcade07IC}.
Another interesting challenge is to see how
sequent calculus can be used to study the more difficult problem of
solving intruder deduction constraints under richer intruder models, 
e.g., like those studied in  \cite{Comon-Lundh03LICS,Chevalier03LICS,Delaune06ICALP}. 
An immediate avenue for future work is to prove the same results as in Section~\ref{sec:constraint},
in particular, the transformation to solved forms,
but for the intruder model with blind signatures. 

It may be of proof theoretic interest to study the exact complexity
of the cut elimination procedure and the translation from natural deduction
to sequent calculus, although these results are not needed in 
establishing the complexity results for the intruder deduction problem.
We leave the complete study of the complexity results for these derivation
transformations to future work. 

\paragraph{Acknowledgement}
This work has been supported by the Australian Research Council 
Discovery Project DP0880549. The authors thank the anonymous referees
of an earlier draft for their helpful comments.

\end{document}